\documentclass[lettersize,journal]{IEEEtran}
\usepackage{cite}
\usepackage{makecell}
\usepackage{amsmath,amssymb,amsfonts}
\usepackage{hyperref}
\usepackage{algorithmic}
\usepackage{graphicx}
\usepackage{xcolor}
\usepackage{stfloats}
\usepackage{tcolorbox}
\usepackage{multirow}
\usepackage{flushend}
\usepackage{subcaption}
\usepackage{algorithmic}
\usepackage{array}
\usepackage{textcomp}
\usepackage{url}
\usepackage{verbatim}
\def\BibTeX{{\rm B\kern-.05em{\sc i\kern-.025em b}\kern-.08em
    T\kern-.1667em\lower.7ex\hbox{E}\kern-.125emX}}
\usepackage{balance}
\begin{document}

\title{On the Ambiguity Function of OFDM-based ISAC Signals Under Non-Ideal Power Amplifiers}
\author{\IEEEauthorblockA{Eya Gourar\IEEEauthorrefmark{1}, Yahia Medjahdi\IEEEauthorrefmark{1}, Laurent Clavier\IEEEauthorrefmark{1}\IEEEauthorrefmark{3}, Abdul Karim Gizzini\IEEEauthorrefmark{2}, Patrick Sondi\IEEEauthorrefmark{1}
}\\
\IEEEauthorblockA{\IEEEauthorrefmark{1} IMT Nord Europe, Institut Mines T\'el\'ecom, Center for Digital Systems, F-59653 Villeneuve d’Ascq, France\\
\IEEEauthorrefmark{3} IEMN, UMR CNRS 8520, University of Lille, France\\
\IEEEauthorrefmark{2} University of Paris-Est Créteil (UPEC), LISSI/TincNET, F-94400, Vitry-sur-Seine, France\\
Email: \ eya.gourar@imt-nord-europe.fr}
}
\maketitle
\begin{abstract}
Integrated Sensing and Communications (ISAC) has garnered significant attention as a promising technology for next-generation wireless and vehicular communications. Among candidate waveforms, Orthogonal Frequency Division Multiplexing (OFDM) has been extensively investigated over the past decade for its robustness against frequency-selective fading and its favorable ranging performance. However, the waveform’s sensing and communication (S\&C) performance depends strongly on the modulation scheme; while variable-amplitude constellations such as quadrature amplitude (QAM) are more efficient for communication, constant-modulus modulations such as phase shift keying (PSK) are more suitable for sensing. Yet, it remains unclear whether these findings persist under power amplifier (PA) nonlinearity. Because OFDM signals exhibit a high peak-to-average power ratio (PAPR), they require highly linear PAs to avoid distortion, which conflicts with radar requirements, where high transmit power is always beneficial for sensing. In this work, we analyze the effect of PA-induced distortions on the sensing task for PSK and QAM constellations. By introducing the Signal-to-Distortion Ratio ($\text{SDR}$), we examine the extent of the distortion limitation on the ranging task. We complement simulation results with a theoretical characterization of the ambiguity function (AF), thereby explicitly demonstrating how distortion artifacts manifest in the zero-Doppler sidelobes (i.e, ranging sidelobes) and the zero-delay sidelobes.
Simulations show that PA distortions impose a palpable performance ceiling for both constellations, reshape the AF, and reduce detection probability, diminishing the theoretical advantage of unimodular signaling and further compromising the OFDM sensing performance with non-uniform envelope signals.
\end{abstract}
\section{Introduction}
\IEEEPARstart{I}{ntegrated} Sensing and Communication (ISAC) is envisioned as a key technology for sixth-generation (6G) networks. Traditionally, communication systems have been implemented on frequencies different from those used by sensing systems, with separate frontends. Moreover, the waveforms used for each of these systems are optimized for the associated function. In the world of 6G communication, which aims for greater spectral efficiency, allowing most services to be implemented via locally available frequencies on both the infrastructure and terminal side, the prospect of using a waveform optimized for both sensing and communication (S\&C) on the same frequency bands is a key research topic.

Such mutualization of sensing and communication functions into a single component offers several advantages across a wide range of 6G applications, including the V2X (Vehicle-to-Everything) communications, with simplified architectures and relatively low-cost hardware \cite{dapa2023vehicular,tong2025integrated,dapa2025parametrizations}. Within the ISAC paradigm, a common strategy to achieve this integration is through mixed modulation of existing signals, exploiting degrees of freedom in multiple dimensions to realize the S\&C tasks \cite{ma2021spatial}. One approach is to utilize radar-centric signals for communication purposes, such
as the Frequency-Modulated Continuous Wave (FMCW) signals that are widely adopted for autonomous vehicles \cite{wang2018joint,temiz2023radar,muja2024real}. However, the achievable data rate is insufficient to meet the requirements anticipated for future V2X applications \cite{dapa2023vehicular}.

\begin{figure*}[htp]
  \centering
  \includegraphics[width=0.8\linewidth]{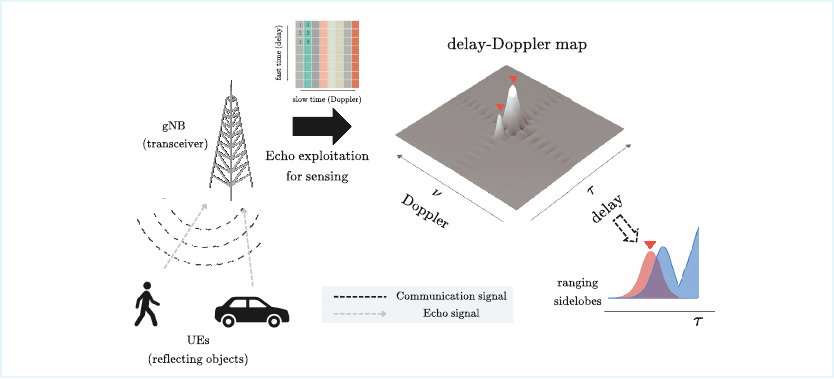}
  \caption{\centering Illustration of communication-centric ISAC and Ambiguity function overlap.}
\label{schema2}
\end{figure*}

In contrast to communication-centric ISAC systems, which are primarily designed for data transmission, but can repurpose their waveforms for sensing. Regarding the sensing task with communication waveforms, the detection of targets depends heavily on their ambiguity functions (AFs) under random data signaling \cite{yang2020dual}. In particular, the sidelobe level indicates the extent of interference from adjacent targets, a factor that is crucial for multi-target sensing in 6G, as illustrated in Fig.~\ref{schema2}, where a base station (e.g., gNB) simultaneously transmits data to user equipments (UEs) and exploits reflected echoes to estimate the reflecting objects' delay-Doppler profile/map. The delay-Doppler map inset shows the potential AF overlap between adjacent targets, motivating AF optimization for accurate multi-target detection.  Building on this, the application of Orthogonal Frequency-Division Multiplexing (OFDM) for sensing purpose seems a natural answer. In \cite{liu2024ofdm}, the authors demonstrated that for quadrature amplitude modulation (QAM) and phase shift keying (PSK) constellations, the conventional OFDM is globally optimal for achieving the lowest average zero-Doppler sidelobes (i.e, ranging sidelobes) level under random signaling. 

Indeed, OFDM radar technology has been extensively investigated for vehicular communication \cite{dokhanchi2018ofdm,dapa2023vehicular,nataraja2024integrated,dapa2025parametrizations,knill2021coded}. Not only does it fulfill the need for high throughput communication between vehicles in multipath fading environments, but it also leverages its inherent ranging capabilities to acquire precise information about the surrounding environment. However, a fundamental challenge lies in the choice of data symbol constellations, since optimal constellations differ for each of the S\&C tasks; a Gaussian distributed constellation maximizes the communications performance for an additive white Gaussian noise (AWGN) channel, whereas constant modulus constellations maximize sensing performance \cite{salman2024sensing,keskin2025fundamental}.
To counter this, many studies have proposed using constellation shaping (probabilistic constellation shaping (PCS) and Geometric shaping (GS)) to generate non-uniform QAM distributions, which improve sensing by narrowing the sensing performance gap between QAM and PSK, while preserving the waveform's randomness (see for example \cite{du2023probabilistic,xu2024experimental,yang2024constellation}) .
However, these studies assumed a linear system, which limits the ability to fully characterize the sensing performance in practical scenarios.
For OFDM in particular, its large Peak-to-Average Power Ratio (PAPR) is a crippling limitation, pushing non-linear power amplifiers (PAs), although realistic, to operate in the highly non-linear region, which causes waveform distortions.

Several works emphasize that such distortions can significantly degrade the sensing performance \cite{feng2024analysis,ismail2024robustness,akca2024integrated,wymeersch2025cross,he2024nonlinear}. In the context of communication-centric ISAC systems, \cite{ismail2024robustness} shows a comparison of the radar performance of different communication waveforms with QAM constellations using both matched filtering and data division receivers, considering a cubic model PA in a monostatic ISAC system. Similarly, \cite{akca2024integrated} investigates the use of PA-distorted signals as matched filters and analyzes their impact on the range-Doppler map with QAM constellations. To mitigate these nonlinearity effects, numerous PAPR-aware waveform design techniques have been proposed (see for example \cite{varshney2023low,bazzi2023integrated,rexhepi2024tone}). However, many existing approaches struggle to achieve low-PAPR waveforms without introducing some degree of signal distortion. \\ As an alternative, \cite{salman2024sensing} proposes using dedicated constant-modulus sensing signals or sensing \textit{"pilots"} to achieve superior sensing performance (e.g., echo SNR) by optimizing the powers of the communication and sensing OFDM signals with a reduced complexity. Although these findings help enrich the ongoing discourse regarding the most effective communication signal for realizing the sensing task, the communications rate might be compromised due to the replacement of a portion of data by pilots  \cite{salman2024sensing,keskin2025fundamental} or the usage of whole unimodular sequences \cite{varshney2023low}. Moreover, in communciation-centric scenarios, the signal's PAPR often remains moderately high \cite{salman2024sensing,gourar2025analysis}, and PA-induced nonlinearities continue to pose a challenge.

Hence, several fundamental questions remain open: Does constant-modulus signaling (e.g., PSK) continue to provide tangible benefits over non-constant-modulus signaling (e.g., QAM) when nonlinearities are present? To what extent does the choice of modulation impact the sensing performance of OFDM radars in such cases? Furthermore, do OFDM signals retain the ranging sidelobe advantages over other standardized waveforms in practical scenarios? In response, this work addresses these questions by combining a theoretical characterization of the PA-induced distortion and the resulting AF with simulations, showing how distortion artifacts affect both the ranging sidelobes and the zero-delay sidelobes, which, to the best of our knowledge, have not been previously explored for communication-centric waveforms. \\

The main contributions are summarized as follows:

\begin{itemize}
    \item We study the effect of non-ideal PAs on the target detection in OFDM, and we predict a performance ceiling by introducing the Signal-to-Distortion Ratio ($\text{SDR}$) and the effective Signal-to-Noise Ratio ($\text{SNR}_{\text{eff}}$).
    \item We derive two generic expressions of the AF of a non-ideally amplified signal by (1) exploiting the Bussgang theorem and (2) conditioning on the signal’s saturation likelihood. We then obtain the expressions for the zero-Doppler and zero-delay cuts to study the sidelobe performance. We define the expectation of the integrated sidelobe level (EISL) as well as its ratio to the mainlobe energy (EISLR) as performance metrics for ranging, and we show that PA-induced distortions alter the AF shape of the OFDM signal, regardless of the model.
    \item We compare the ranging performance of OFDM with other standardized communication waveforms (Code Division Multiple Access (CDMA) and Single Carrier (SC)) under PA distortions. That is to say, the OFDM waveform still achieves the lowest ranging sidelobes.
    \item We investigate the implications of the constellation choice (i.e., QAM and PSK) on ranging performance in terms of sidelobe levels, EISL$\cdot$R, and probability of detection with simulations across multiple PA configurations with and without the inclusion of a CP. An intuitive discussion of the analytical results further supports the study.
\end{itemize}
\begin{figure*}
  \centering
  \includegraphics[width=0.65\linewidth]{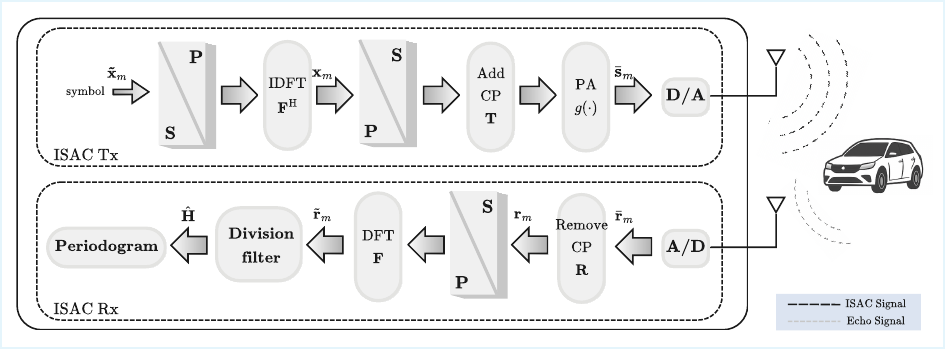}
  \caption{\centering Block diagram of monostatic ISAC transceiver, where the signal reflected by the target is received by the receiving end for sensing purposes.}
\label{schema}
\end{figure*}

In order to ensure a clear interpretation of the proposal, the following notations are used throughout this work:
\begin{itemize}
    \item Column vectors, matrices, and scalars are denoted by boldface lowercase, boldface uppercase letters, and italics, respectively.
    \item The superscripts $(\cdot)^\text{H}$ and $(\cdot)^\text{T}$ denote the complex conjugate transpose and the transpose, respectively.
    \item $x(n)$ denotes the $n^\text{th}$ element of a vector $\mathbf{x}$. 
    \item $X(n,m)$ denotes the element at the intersection of the $n^\text{th}$ line and $m^\text{th}$ column of a matrix $\mathbf{X}$.
    \item The time-domain signal $\mathbf{x}$ after cyclic prefix (CP) insertion is denoted as $\bar{\mathbf{x}}$, and $\tilde{\mathbf{x}}$ denotes the frequency-domain representation.
    \item $\mathbf{I}_N$ denotes the ($N\times N$) identity matrix.
    \item $\mathbf{0}$ and $\mathbf{1}$ denote the ones and zeros matrices of appropriate dimension, respectively. \item The notations $|\mathbf{x}|$ and $\arg(\mathbf{x})$ denote the element-wise modulus and argument of the entries of a complex vector $\mathbf{x}$, respectively. 
    \item $\oslash$ denotes the element-wise division.
    \item $\Re\{\cdot\}$ denotes the real part of a complex number. 
    \item $\mathbb{P}(\cdot)$ is the probability operator.
    \item The notation $\text{diag}(\mathbf{a})$ denotes the diagonal matrix obtained by placing the entries of $\mathbf{a}$ on its main diagonal.
    \item $\delta(x)$ is the unit impulse function defined by $\delta(x) = 1$ iff $ x = 0$.
\end{itemize}

The paper is organized as follows: Section \ref{section1} introduces the system model for the sensing problem. Section \ref{section2} analytically characterizes the AF of the PA-distorted signal. Finally, section \ref{section3} presents the simulation results, prior to our conclusions.

\section{System Model}\label{section1}
\subsection{ISAC Signal Model}
In an OFDM-based ISAC system, a single OFDM symbol vector consists of $N$ complex values representing the symbols that will be transmitted on the communication subcarriers, each occupying a duration $T$ in the time domain. A transmitted frame contains $M$ OFDM symbols. To simplify the notations, the processing at the transmitter is described for the $m^\text{th}$ OFDM symbol in the frame. The time-domain (TD) OFDM signal (denoted by $\mathbf{x}_m$) is generated taking the inverse discrete Fourier transform (IDFT) of the OFDM symbol $\mathbf{\tilde{x}}_m$, i.e., $\mathbf{x}_m =\mathbf{F}_\text{N}^\text{H}\mathbf{\tilde{x}}_m$, where $\mathbf{F}_\text{N}\in \mathbb{C}^{N\times N}$ denotes the normalized DFT matrix of size $N$ whose $(n,n')^\text{th}$ element is defined as follows
\begin{equation}
    \mathbf{F}_\text{N}(n,n')=\frac {1}{\sqrt {N}}e^{-j \frac {2\pi }{N}n\,n'} \,(\,n,n'=0, {\dots }, N-1)
\end{equation}

A mono-static radar receiver that is colocated with the transmitter and has full knowledge of the transmitted signal $\mathbf{x}_m$, is assumed for sensing. Fig.~\ref{schema} shows the block diagram of the transmitter and the radar receiver.
To recover the transmitted signal at the communication receiver without inter-symbol interference (IBI), a CP of length $L$ is added to the transmit symbol $\mathbf{x}_m$. The resulting signal before amplification is denoted $\bar{\mathbf{x}}_m=\mathbf{T}\mathbf{x} _m\in \mathbb{C}^{(L+N) \times 1}$, where $\mathbf{T}=[\mathbf{L},\mathbf{I}_N]^\text{T} \in \mathbb{C}^{(L+N) \times N}$ is the CP addition matrix, $\mathbf{L}$ contains the last $L$ rows of $\mathbf{I}_N$.
The TD signal $\mathbf{\bar{s}}$ at the output of a non-ideal PA can be expressed as
\begin{equation}\mathbf{\bar{s}}_m=g(\mathbf{\bar{x}}_m)=G \mathbf{\bar{x}}_m+\mathbf{\bar{d}}_m = \textbf{T}(G\textbf{x}_m + \textbf{d}_m).
\label{sxd}\end{equation}

In \eqref{sxd}, $G$ is a fixed input complex scaling factor that models the amplifier’s linear gain, and $\mathbf{\bar{d}}_m$ is a random variable representing the PA-induced distortion.

We assume a single receive antenna and a sampled multipath channel with $H$ propagation paths, where each path $h$ is described by a complex attenuation factor $b_h$, a propagation delay $\tau_h$, and a Doppler frequency $\nu_h$. We denote the sampling period by $T_s$ (so that $T_s=T/N$) and the integer delay and normalized Doppler are then
\begin{equation}
l_h \triangleq \frac{\tau_h}{T_s}, \qquad k_h \triangleq N \nu_h T_s,
\end{equation}
with $l_h$ assumed integer when $T_s$ is sufficient. Let $p\in\{0,\cdots,N-1\}$ and $l\in\{0,\cdots,N-1\}$ denote the discrete time and delay indices.
The discrete-time sampled impulse response at symbol $m$ is
\begin{equation}
h_m(p) = \sum_{h=1}^{H} b_h \, e^{j 2 \pi \frac{k_h}{N}(L+N)m} \, \delta(p - l_h),
\label{eq:hh_timevarying}
\end{equation}

We define the two $(N+L)\times(N+L)$ Toeplitz matrices $Q_m(n,n')=h_m(n-n')$ representing the convolution of the current transmit symbol, and $J_m(n,n')=h_m(L+N + n-n')$ representing the IBI caused by the previous transmit symbol $\mathbf{s}_{m-1}$. We also denote $\bar{\mathbf{z}}_m$ the additive noise such that $\bar{z}_m(p) \sim \mathcal{CN}(0, \sigma^2_z)$, then the $m^\text{th}$ received vector at the radar receiver is given by \cite{ismail2024robustness}
\begin{equation}
\mathbf{\bar{y}}_m =  \mathbf{Q}_m\bar{\mathbf{s}}_m + \mathbf{J}_m\bar{\mathbf{s}}_{m-1} + \bar{\mathbf{z}}_m \in \mathbb{C}^{(L+N)\times 1}.
\label{CPr}
\end{equation}

\subsection{Non-Ideal PA Model}
A PA model or a real measured one can be described by its transfer function $g(\cdot)$.

\subsubsection{Soft Envelope Limiter (SEL)}
a simple clipping PA model is considered. The input-output relationship is given by
\begin{equation}
\begin{aligned}
\bar{s}_m(p) &= g(\bar{x}_m(p)) = \begin{cases}G \bar{x}_m(p)&, G|\bar{x}_m(p)| \leq V_{sat} \\ V_{sat} e^{j\arg{(\bar{x}_m(p))}}&, \text {otherwise} \end{cases} ,
\end{aligned}
\end{equation}
where $V_{sat}$ is a SEL parameter indicating the saturation threshold.
The PA output distortion in \eqref{sxd} can be written as
\begin{equation}
\bar{d}_m(p) 
= \begin{cases} 0&, G|\bar{x}_m(p)| \leq V_{sat}\\ V_{sat} e^{j\arg{\bar{x}_m(p))}} - \bar{x}_m(p)&,  \text {otherwise} \end{cases} \;.
\label{s1SEL}
\end{equation}
In practice, to mitigate the impacts of the non-linear distortion, the PA operates at an input back-off (IBO) from a given saturation level. In the literature, there have been many definitions of the IBO \cite{bouhadda2014theoretical}. In this work, we adopt the following definition 
\begin{equation}
    \text{IBO} =  \frac{\text{P}_{\text{1dB}}}{\sigma ^{2}}  ,
\end{equation}

where $\sigma^2=\mathbb{E}[|\mathbf{\bar{x}}_m|^2]$ is the mean input signal power, and $\text{P}_{\text{1dB}}$ is the input power at the 1 dB compression point. Therefore, the input signal is multiplied beforehand by the back-off coefficient $\alpha \in ]0,1]$ 
\vspace{-0.1cm}
\begin{equation}
\alpha = \frac{1}{\sigma}\sqrt{\frac{\text{P}_{\text{1dB}}}{ \text{IBO}}} .
\end{equation}

\subsubsection{Bussgang Decomposition}
It is worth noting that, as in conventional OFDM systems, when the number of subcarriers $N$ is large, the TD signal results from the sum of independent modulated subcarriers. According to the central limit theorem, $\mathbf{\bar{x}}_m$ can therefore be approximated as i.i.d.\ complex Gaussian distributed \cite{ismail2024robustness}. Based on the Bussgang theorem, the output of the non-ideal PA circuit can be expressed in terms of a linear scale parameter $\kappa$ of the input signal and a nonlinear distortion $\mathbf{\bar{d}}_m$ which is uncorrelated with the input signal. In this case, the output of the SEL is approximated by \cite{bouhadda2014theoretical}
\begin{equation} \mathbf{\bar{s}}_m = \kappa \mathbf{\bar{x}}_m + \mathbf{\bar{d}}_m ,\label{s1}\end{equation} 

where $\kappa$ is calculated as follows 

\begin{equation}
\kappa =\frac{\mathbb{E}[\mathbf{\bar{x}}_m^\ast \mathbf{\bar{s}}_m]}{\mathbb{E}[|\mathbf{\bar{x}}_m|^2]}.
\end{equation} 
and $\mathbb{E}[|\mathbf{\bar{d}}_m|^2]=\sigma_d^2$ is the variance of the distortion $\mathbf{\bar{d}}_m$,

\subsection{2D-FFT-Based ISAC Processing}

For target detection, we generate a 2D periodogram of the targets by applying a 2D-FFT to the received signal. Prior to that, the received signal is filtered using a known reference signal \cite{braun2014ofdm}. The monostatic case utilizes the transmit frequency-domain (FD) signal, which is fully known at the ISAC transceiver. Therefore, the transmitted and received signals should be expressed in the frequency domain. The received signal in \eqref{CPr} is first multiplied by the CP removal matrix $\mathbf{V} = \begin{bmatrix} \mathbf{0}_{N \times L} & \mathbf{I}_N\end{bmatrix}$, yielding
\begin{equation}
\mathbf{y}_m = \mathbf{V} \mathbf{\bar{y}}_m =  \mathbf{V} \big(\mathbf{Q}_m \, \bar{\mathbf{s}}_m + \mathbf{J} \bar{\mathbf{s}}_{m-1} + \bar{\mathbf{z}}_m \big) \in \mathbb{C}^{N \times 1}.
\label{r}
\end{equation}
Adding that $\bar{\mathbf{d}}_m= \mathbf{T}\mathbf{d}_m $, the second term in \eqref{r} can be expressed as $\mathbf{V}\mathbf{J}_m\mathbf{T}(G\mathbf{x}_{m-1} + \mathbf{d}_{m-1})$, which vanishes as $\mathbf{V}\mathbf{J}_m\mathbf{T} = 0$, given that the CP length is at least equal to the channel maximum delay. Next, substituing \eqref{sxd} in \eqref{r}, and noting $\mathbf{z}_m= \mathbf{V}\bar{\mathbf{z}}_m$, the received signal can be expressed as
\begin{equation}
    \mathbf{y}_m 
    = \mathbf{V} \mathbf{Q}_m \mathbf{T}\big( G \mathbf{x}_m+\mathbf{d}_m\big) + \mathbf{z}_m  = \mathbf{H}_m (G \mathbf{x}_m+\mathbf{d}_m) 
    + \mathbf{z}_m,
    \label{eq:r1}
\end{equation}
where $\mathbf{H}_m \in \mathbb{C}^{N \times N}$ is a circulant matrix due to the CP removal, which implies that $\mathbf{H}=\mathbf{F}_\text{N}^\text{H} \, \Lambda_m \, \mathbf{F}_\text{N}$, where the diagonal matrix $\boldsymbol{\Lambda}_m = \mathrm{diag}(\tilde{\mathbf{h}}_m)$ contains the FD channel response $\tilde{\mathbf{h}}_m$. 
Therefore, we can write the FD received signal as follows 
\begin{equation}
    \tilde{\mathbf{y}}_m 
    = \mathbf{F}_\text{N}\mathbf{y}_m =  \Lambda_m \, (G \tilde{\mathbf{x}}_m+\tilde{\mathbf{d}}_m) + \tilde{\mathbf{z}}_m ,
    \label{eq:r_freq}
\end{equation}
where $\tilde{\mathbf{d}}_m\in \mathbb{C}^{N\times 1}$ and $\tilde{\mathbf{z}}_m\in \mathbb{C}^{N\times 1}$ are the PA distortion and additive noise in the frequency domain, and the equivalence between equations \eqref{eq:r1} and \eqref{eq:r_freq} follows from the linearity of the DFT and the fact that $\tilde{\mathbf{d}}_m=\mathbf{F}_\text{N} \, \mathbf{d}_m$ and $\tilde{\mathbf{z}}_m=\mathbf{F}_\text{N} \, \mathbf{z}_m$.
To extract the channel frequency response, we filter the received signal using the reference signal $\tilde{\mathbf{x}}_m $, i.e., the original transmitted signal before PA distortions, yielding
\begin{equation}
    \mathbf{\hat{h}}_m = \tilde{\mathbf{y}}_m\oslash\tilde{\mathbf{x}}_m
    =  \Lambda_m \, (G \, \mathbf{1}_{N\times 1} + \tilde{\mathbf{d}}_m\oslash\tilde{\mathbf{x}}_m) + \tilde{\mathbf{z}}_m\oslash\tilde{\mathbf{x}}_m \,.
    \label{hatH}
\end{equation}
By applying the data division filter in \eqref{hatH} to each received block with its corresponding symbol vector in the transmitted frame, we extract the channel coefficients, which are stacked into a matrix  $\mathbf{\hat{H}}$ from which the periodogram $\textbf{Per} \in \mathbb{C}^{N_\text{Per}\times M_\text{Per}}$ of the targets is constructed as follows \cite{braun2014ofdm}
\small
\begin{align}
\mathrm{Per}(l,k) 
&= \frac{1}{NM} \left| 
   \sum_{n=0}^{N_{\text{Per}}-1} 
   \left( \sum_{m=0}^{M_{\text{Per}}-1} 
   \hat{H}(n,m)  e^{-j 2 \pi \frac{km}{M_{\text{Per}}}} \right) 
   e^{j 2 \pi \frac{ln}{N_{\text{Per}}}}
   \right|^2,
\end{align}
\normalsize
where $l= 0,\cdots, N_{\text{Per}}-1 $ and $k=\frac{-M_{\text{Per}}}{2}, \cdots ,\frac{M_{\text{Per}}}{2}-1$.
We fix $N_{\text{Per}} \geq N$ and $M_{\text{Per}} \geq M$ as the number of supporting points of the discrete periodogram on the delay and Doppler axis, respectively. If $N_{\text{Per}} > N$ and $M_{\text{Per}} > M$, zero-padding is used to increase the dimensions of $\mathbf{\hat{H}}$ to $N_{\text{Per}}$ and $M_{\text{Per}}$.

The division filter can alter the statistical properties of the noise across subcarriers and OFDM symbols, resulting in an increased sidelobe level in the periodogram floor. Similarly, PA-induced distortion is expected to elevate the sidelobe floor of targets. This elevation can make weak target reflections indistinguishable from the sidelobe floor, leading to missed targets and false detections.
While increasing the signal power limits the noise impact, the PA-induced distortion is signal-dependent and cannot be mitigated in a similar manner. To quantify this limitation, we introduce the Signal-to-Distortion Ratio (SDR), 
\begin{equation}
\text{SDR} =\frac{\mathbb{E}[ G^2|\mathbf{x}_m|^2]}{\mathbb{E}[|\mathbf{d}_m|^2]}= \frac{ G^2\alpha^2\sigma^2 }{\sigma_d^2} = \frac{ G^2\text{P}_{\text{1dB}} }{\text{IBO} \,\sigma_d^2} .
\label{SDRSEL}
\end{equation}
As IBO decreases, the input power increases and the PA moves out of its linear operating region, distortion becomes more significant, causing the SDR to either plateau or decline, suggesting a performance ceiling. Increasing the power of the input signal (or the input Signal-to-Noise Ratio ($\text{SNR}_{\text{0}} = \frac{G^2\alpha^2\sigma^2}{\sigma_z^2 }$)) does not mitigate the distortion impact, since not all input power is being used. We introduce then the system's effective SNR ($\text{SNR}_{\text{eff}}$), which can be expressed as 
\begin{equation}
\text{SNR}_{\text{eff}} = \frac{G^2\alpha^2\sigma^2}{\sigma_z^2 + \sigma_d^2 }
\end{equation}
is the SNR that a system under PA distortion needs to achieve the same performance as a linear system. It can be expressed differently as follows 
\begin{equation}
\text{SNR}_{\text{eff}} = \frac{\text{SNR}_{\text{0}}}{1 +   \frac{\text{SNR}_{\text{0}}}{\text{SDR}}} .
\label{SNReff}
\end{equation}
In \eqref{SNReff}, at a very high $\text{SNR}_{\text{0}}$ ($\xrightarrow{}+\infty$), the effective SNR converges to the SDR, meaning that the PA distortion is a limiting factor even at high power levels.

Although target detection is performed on the periodogram and not directly on  $\hat{\mathbf{h}}_m$, its signal-to-noise ratio $\text{SNR}_{\text{Per}}$ can nonetheless be expressed in terms of the SNR that is just before the signal processing, in $\tilde{\mathbf{y}}_m$ (Consequently $\mathbf{\hat{h}}_m$ since the division with $\tilde{\mathbf{x}}_m$ does not change the SNR) as $\text{SNR}_{\text{Per}}= \text{SNR} + 10 \log_{10}(N_{\text{Per}} M_{\text{Per}})$, on the logarithmic scale \cite{braun2014ofdm}.

\section{Ambiguity function of non-ideally amplified Signals}\label{section2}
The ambiguity function (AF) describes the ability of a waveform to precisely determine the time delay $\tau$ and Doppler frequency shift $\nu$ that are related to a target. The AF can be characterized as either the linear or periodic matched filter of the OFDM signal, depending on whether a cyclic prefix is included or not. Still, in both cases, the per-symbol analysis suffices to capture the waveform’s delay-Doppler representation. Therefore, the symbol index $m$ is omitted for clarity. Following our CP-inclusive case, the discretized periodic AF (PAF) is defined as follows 
\begin{equation}
A(l,k) = \frac{1}{\sqrt{N}} \sum _{p=0}^{N-1} s(p) s^\ast (\small(p-l\small)\text{mod}\small(N\small) \big) e^{-j2\pi  \frac{kp}{N}},
\label{AF_discrete}
\end{equation}
where the spacing of the sampling period is $T_s$ in the time axis and $1/(K T_s)$ in the Doppler shift axis where $K$ is the number of frequency grids. Therefore $l=\tau/T_s$ and $k=\nu(K T_s)$.

We define the integrated sidelobe level (ISL) as the energy distributed in the sidelobes i.e., $\text{ISL} = \sum^{N-1}_{l=1-N,l \neq 0}|A(l,0)|^2$. It is an important metric in dense target scenarios and when distributed clutter is present.
Due to the random nature of the ISAC signal and the AF, we define the expectation of the sidelobe level as a performance metric.
\begin{equation}
  \text{EISL} =  \sum^{N-1}_{l=1-N,l \neq 0} \mathbb{E}[|A(l,0)|^2]  .
\end{equation}
We further define the EISLR as the average ratio of the energy integrated over sidelobes to the total mainlobe energy $|A(0,0)|^2$, as 
\begin{equation}
  \text{EISLR} =  \frac{1}{|A(0,0)|^2}\sum^{N-1}_{l=1-N,l \neq 0} \mathbb{E}[|A(l,0)|^2]  .
\end{equation}
Lower EISLR values indicate that the overall sidelobes energy relative to the mainlobe energy is small, which contributes to improved target detection.

In practice, the signal is subject to PA-induced distortions prior to transmission. These distortions surely modify the structure of the transmitted waveform, and consequently alter its AF. To characterize these effects analytically, we extend the general AF expression in \eqref{AF_discrete} to the case of non-ideally amplified signals. 

In the following, we extend the AF analysis under the Bussgang approximation to investigate how PA nonlinearities generally alter the ranging performance. We then narrow the study to the SEL model to further quantify the distortion effect on both the zero-Doppler and zero-delay cuts, while examining the implications of the constellation of the data symbols.

\subsection{Ambiguity Function under Bussgang Approximation}\label{section2:BussAF}

To account for PA-induced distortions, we analyze the AF using the Bussgang approximation. The following expression formalizes this,
\small
\begin{equation}
\begin{aligned}
    A(l,k) &= \kappa^2 A_x(l,k) + \kappa A_{x,d}(l,k)  + \kappa^\ast A_{d,x}(l,k) + A_d(l,k),
\end{aligned}
\end{equation}
\normalsize
where the self-AFs and cross-AFs are defined, respectively, as
\begin{align*}
    A_{u}(l,k) =  \frac{1}{\sqrt{N}} \sum _{p=0}^{N-1} u(p) {u^\ast\big((p-l)\!\!\mod N\big)} e^{-j2\pi kp/N}, \\
    A_{u,v}(l,k) =  \frac{1}{\sqrt{N}} \sum _{p=0}^{N-1} u(p) {v^\ast\big((p-l)\!\!\mod N\big)} e^{-j2\pi kp/N}.
\end{align*}
\textit{Proposition 1.} \textit{The squared PAF under Bussgang decomposition can be written as in} \eqref{AFBussgang}.
\begin{figure*}[htp]
\begin{equation}
\begin{aligned}
    |A(l,k)|^2 &= |\kappa|^4 |A_x(l,k)|^2 + |\kappa|^2\left( |A_{x,d}(l,k)|^2 + |A_{d,x}(l,k)|^2 \right) + |A_d(l,k)|^2 \\
    &\quad + 2\,\Re\!\big\{\,|\kappa|^{2}\kappa\,A_xA_{x,d}^{\ast}
+ \kappa^{3}\,A_xA_{d,x}^{\ast}
+ \kappa^{2}\,A_xA_d^{\ast} \\
& \qquad\qquad + \kappa^{2}\,A_{x,d}A_{d,x}^{\ast}
+\kappa\,A_{x,d}A_d^{\ast}
+ \kappa^{\ast}\,A_{d,x}A_d^{\ast}\big\}\,.
\end{aligned} 
\label{AFBussgang}
\end{equation}
\noindent\rule{\textwidth}{0.2pt}
\end{figure*}

It is expressed as the sum of the AF of the input signal $\mathbf{x}$, the AF of the distortion $\mathbf{d}$, and other distortion components expressed in terms of $A_{x,d}(l,0)$, $A_{d,x}(l,0)$, $A_{x,d}^\ast(l,0)$ and $A_{d,x}^\ast(l,0)$. It clearly reveals that PA nonlinearities affect the overall AF time and frequency structure in a broad sense, regardless of the PA model specification.\\

\noindent \textit{Proposition 2.} \textit{The squared zero-Doppler cut of the PAF under Bussgang decomposition reduces to}
\begin{align}
|A(l, 0)|^2 &= |\kappa|^4 |A_x(l, 0)|^2 + |A_d(l, 0)|^2 \nonumber \\
&\quad +  2\Re\{\kappa^2 A_x(l,0) A_d^\ast(l,0)\} \, \, \,.
\label{A22}
\end{align}

\textit{Proof:} It can be readily shown that for $k=0$, the input signal $\mathbf{x}$ and the distortion $\mathbf{d}$ are uncorrelated, so the cross terms $A_{x,d}(l,0)$ and $A_{d,x}(l,0)$ vanish.\hfill\IEEEQED \\

Ideally, to resolve weak targets in the presence of interference from adjacent targets, the zero-Doppler cut should have zero sidelobe energy. This is achieved with unimodular transmit symbols (i.e., $|\tilde{x}(p)|=1$), such as PSK, resulting in a perfect thumbtack PAF with zero ISL \cite{varshney2023low}. Under non-linear transmission as in \eqref{s1}, the sidelobe energy becomes
\begin{equation}
\text{ISL} = \sum_{\substack{l=1-N,\\ l\neq 0}}^{N-1} \left( |A_d(l, 0)|^2  + 2\Re\{\kappa^2 A_x(l,0) A_d^\ast(l,0)\} \right) > 0.
\end{equation}
Thus, even for perfectly designed unimodular transmit sequences with ideal thumbtack PAFs, non-linear PAs inevitably induce sidelobe regrowth through the distortion terms, leading to degraded ranging performance.\\

\textit{Corollary 1.} \textit{The expected integrated sidelobe level of the zero-Doppler cut under Bussgang decomposition is}\label{coroll1}
\small
\begin{equation}
\begin{aligned}
\text{EISL} &= \frac{1}{N} \sum^{N-1}_{\substack{l=1-N,\\ l\neq 0}} \sum^{N-1}_{p=0} |\kappa|^4\mathbb{E}\left[  |x(p)|^2 |x(p-l)|^2  \right] \\ & \quad+ \mathbb{E}\left[  |d(p)|^2 |d(p-l)|^2  \right]\\  & \quad+ \frac{2\kappa^2}{N}\Re\{\sum^{N-1}_{\substack{p=0,\\ p'= 0}} \mathbb{E}\big[  x(p)x^\ast(p-l) d^\ast(p')d(p'-l) \big]\} \\ &= (2N-2)\left(|\kappa|^4 \sigma^4 + \sigma_d^4 \right) .
 \end{aligned}
\label{Buss_EISL}
\end{equation}
\normalsize
\textit{Proof:} Please refer to Appendix \ref{A}.\\
This result shows that under PA distortions, the EISL scales linearly with the number of symbol samples $N$.\\

\textit{Corollary 2.} \textit{The expected mainlobe level of the PAF under Bussgang decomposition can be expressed as}
\small
\begin{equation}
\mathbb{E}\left[ |A(0, 0)|^2 \right] = |\kappa|^42\sigma^4 + \mathbb{E}\left[  |d(p)|^4 \right]  + 2 \kappa^2N\sigma^2\sigma^2_d.
\label{Buss_EISLR}
\end{equation}
\normalsize
\textit{Proof:} Please refer to Appendix \ref{A}.\\
This indicates that the mainlobe also grows linearly with $N$ under PA distortions, but grows faster than each sidelobe, which contributes only a fraction of the EISL.\\

\textit{Remark 1.} The extension of these results to the aperiodic case is straightforward and remains valid, as the assumptions of i.i.d. TD samples and uncorrelated distortion still hold.
\subsection{Ambiguity Function under the SEL Model}
\label{section2:SEL}

For the purpose of extending the discussion to the case where no CP is applied, we will focus on the aperiodic ambiguity function (AAF). For notation simplicity, we fix $G=1$. We first establish the general expression of the AAF under the SEL model as follows
\small
\begin{equation} \begin{aligned} A(l,k) &= \frac{1}{\sqrt{N}} \sum _{p=0}^{N-1} s(p){s^\ast(p - l )} e^{-j2\pi \frac{kp}{N}} \, \\ &= \frac{1}{\sqrt{N}} \times \big(\mathbb{P}(|x(p)| \leq V_{sat},\, |x(p-l)| \leq V_{sat}) \\& \quad \,\times \sum _{p=0}^{N-1}x(p) x^\ast(p-l) e^{-j2\pi \frac{kp}{N}} \\ & \quad \,+ \mathbb{P}(|x(p)| \leq V_{sat},\, |x(p-l)| > V_{sat}) V_{sat} \\ &\quad \, \times \sum _{p=0}^{N-1} x(p) e^{-j \arg(x(p-l))}e^{-j2\pi \frac{kp}{N}} \\ & \quad \,+ \mathbb{P}(|x(p)| > V_{sat},\, |x(p-l)| \leq V_{sat}) V_{sat} \\ &\quad \,\times \sum _{p=0}^{N-1} x(p-l) e^{j \arg(x(p))} e^{-j2\pi \frac{kp}{N}} \\ & \quad \,+ \mathbb{P}(|x(p)| > V_{sat},\, |x(p-l)| > V_{sat}) V_{sat}^2 \\ & \quad \, \times \sum _{p=0}^{N-1} e^{j (\arg(x(p)) - \arg(x(p-l)))} e^{-j2\pi \frac{kp}{N}} \big) . \end{aligned} \label{SELAF}\end{equation}\normalsize
\noindent 
The AF of the signal at the output of an SEL can be approximated by conditioning on whether the instantaneous magnitudes $|x(p)|$ and $|x(p-l)|$, are clipped by the saturation threshold $V_{sat}$. The first term corresponds to the undistorted elements of the signal, while the remaining terms introduce the distortion-related component.\\

\noindent\textit{Proposition 3.} \textit{ The zero-Doppler cut of the AAF under the SEL model can be expressed as}
\begin{equation} \begin{aligned} A(l, 0) &= \mathbb{P}_<(Y) A_x(l,0)\\ & \quad \,+ \frac{1}{\sqrt{N}} \Big( 2 \big(1 - e^{-Y^2} - \mathbb{P}_<(Y)\big) V_{sat}) \\ & \quad \, \times \sum _{p=0}^{N-1} \big(x(p) e^{-j \arg(x(p-l))} + x(p-l) e^{j \arg(x(p))}\big) \\ & \quad \, + \big(2e^{-Y^2} + \mathbb{P}_<(Y) - 1\big)V_{sat}^2 \\ & \quad \, \times \sum _{p=0}^{N-1} e^{j (\arg(x(p)) - \arg(x(p-l)))} \Big) . \end{aligned} \label{SELACF} \end{equation}
where $Y=\frac{V_{sat}}{\alpha\sigma}$ and $\mathbb{P}_<(Y)$ is defined in the following
\small
\begin{equation*}
\begin{aligned}
\mathbb{P}_<(Y) &=\mathbb{P}(|x(p)| \leq V_{sat},\ |x(p-l)| \leq V_{sat}) \\
&= \begin{cases} 
1 - 2e^{-Y^2} + \frac{1}{2\pi} \int_{-\pi}^{\pi} e^{ -Y^2 \left[ 
1 +  \frac{(1-\rho(l))}
{1 + 2\sqrt{\rho(l)} \sin \theta + \rho(l)} 
\right]} d\theta \\ \qquad\qquad\qquad\qquad\qquad\qquad\qquad\qquad\qquad, 0<l,\\
1 - e^{-Y^2}  \qquad\qquad\qquad\qquad\qquad\qquad\qquad\,, l =0
\end{cases}
\end{aligned}
\label{eq:proba}
\end{equation*}
\normalsize
\textit{Proof:} Please refer to Appendix \ref{C}.

The contribution of each term depends on the joint clipping likelihood of $(x(p),x(p-l))$. As the signal approaches the saturation region (i.e., smaller IBO), the weight of the clipped components represented by the joint clipping probabilities; \small$\mathbb{P}(|x(p)| > V_{sat},\, |x(p-l)| \leq V_{sat}),\mathbb{P}(|x(p)| \leq V_{sat},\, |x(p-l)| > V_{sat})\,$\normalsize and \small$\mathbb{P}(|x(p)| > V_{sat},\, |x(p-l)| > V_{sat})$\normalsize, increases, while that of the undistorted autocorrelation $A_x(l,0)$ corresponding to the non-clipping probability $\mathbb{P}_<(Y)$, decreases. Because clipping constrains amplitude variations, the distortion terms become less random in magnitude, thereby disrupting the original correlation properties of the signal. As a result, the ranging sidelobe levels tend to rise after PA clipping.\\

\noindent\textit{Corollary 3. (The relative sidelobe  rise)}\label{coroll2}:
The constellation statistics determine the extent to which clipping distorts the zero-Doppler cut.
For PSK constellations, which exhibit nearly constant envelopes, even mild clipping can substantially reduce the contribution of $A_x(l,0)$, whose sidelobes are originally well suppressed. This results in a pronounced growth in sidelobe levels.
In contrast, QAM constellations exhibit inherent amplitude variability, leading to moderate clipping even at high IBO values, and, consequently, to more frequent clipping occurrences. However, severe clipping does not necessarily produce a marked increase in sidelobe levels relative to those of $A_x(l,0)$, since QAM signals already exhibit higher ranging sidelobes prior to amplification.\\

\noindent\textit{Remark 2.}\label{remark1}
The parameter $\alpha$ controls how close the signal power is to the saturation region. Higher $\alpha$ (i.e., smaller $Y$) corresponds to higher clipping probability.
Consequently, the sidelobe level of $A(l,0)$ rises with $\alpha$, or as we consider a lower IBO.\\

\noindent\textit{Proposition 4.}
\textit{The expected integrated sidelobe level under the SEL model is expressed as}
\begin{equation}
\begin{aligned}
\text{EISL} &= 
\sum_{\substack{l=1-N,\\ l\neq 0}}^{N-1}
\Bigg(
\sum_i \mathbb{E}\big[|W_i(l)|^2\big]
+ \sum_{i \neq j} \mathbb{E}\big[W_i(l)W_j^\ast(l)\big]
\Bigg),
\end{aligned}
\label{EISLSEL}
\end{equation}
where
\noindent\begin{equation*}
W_1(l)= \mathbb{P}_<(Y) A_x(l,0)
\end{equation*}
\begin{equation*}
W_2(l) = \big(1 - e^{-Y^2} - \mathbb{P}_<(Y)\big) \frac{V_{sat}}{\sqrt{N}}
\sum_{p=0}^{N-1} x(p)e^{-j\arg(x(p-l))}
\end{equation*}
\begin{equation*}
W_3(l) = \big(1 - e^{-Y^2} - \mathbb{P}_<(Y)\big) \frac{V_{sat}}{\sqrt{N}}
\sum_{p=0}^{N-1} x(p-l)e^{j\arg(x(p))}
\end{equation*}
\begin{equation*}
W_4(l) = \big(2e^{-Y^2} + \mathbb{P}_<(Y) - 1\big) \frac{V_{sat}^2}{\sqrt{N}}
\sum_{p=0}^{N-1} e^{j(\arg(x(p))-\arg(x(p-l)))}
\end{equation*}

\noindent
 
Each $W_i(l)$ represents a contribution associated with a specific pairwise clipping case and is weighted by the probability factors $\mathbb{P}_<(Y)$,$\big(1 - e^{-Y^2} - \mathbb{P}_<(Y)\big)$, and $\big(2e^{-Y^2} + \mathbb{P}_<(Y) - 1\big)$. As the signal approaches the saturation region (i.e., for smaller $Y$), the contribution of the undistorted term $W_1(l)$ decreases while the distortion terms $W_2(l), W_3(l)$ and $W_4(l)$ dominate, resulting in the expected growth of the sidelobe energy.\\

\noindent\textit{Proposition 5.}
\textit{The AAF zero-delay cut under the SEL model can be expressed as}
\small
\begin{equation}
\begin{aligned}
A(0,k) &= \frac{1}{\sqrt{N}} \sum_{p=0}^{N-1} |s(p)|^2e^{-j2\pi \frac{kp}{N}} \\
&= \frac{1}{\sqrt{N}} 
\Big(
(1 - e^{-Y^2}) \sum_{p=0}^{N-1}|x(p)|^2 e^{-j2\pi \frac{kp}{N}}
+ V_{sat}^2 e^{-Y^2}\delta(k)
\Big)\,.
\end{aligned}
\end{equation}
\normalsize
\textit{Proof:}  It can be shown that for $(l=0,k\neq0)$, the second and third terms in \eqref{SELAF} are invalid, and only the fully clipped and non-clipped terms remain. The corresponding probabilities thus reduce to the clipping and non-clipping probabilities of the Rayleigh-distributed variable $|x(p)|$, given respectively by $e^{-Y^2}$ and $(1-e^{-Y^2})$.\hfill\IEEEQED
\noindent
 
As nonlinear distortion increases, the term $e^{-Y^2}V_{sat}^2\delta(k)$ (which is only non-zero at $k=0$) becomes dominant, thereby increasing the mainlobe amplitude while flattening the sidelobes of $A(0,k)$.  
Since this effect is primarily amplitude-driven, the zero-delay cut appears nearly identical for PSK and QAM constellations that satisfy zero mean, unit power, and rotational symmetry \cite{liu2024ofdm}.\\

\noindent\textit{Remark 3.}
\textit{(Time–frequency trade-off induced by a nonlinear PA) :} When the signal undergoes amplitude clipping, the instantaneous signal power $ |s(p)|^2 $ becomes nearly constant over time. Its discrete Fourier transform, $\sum_{p=0}^{N-1} |s(p)|^2 e^{-j2\pi kp/N}$, thus concentrates most of its energy around $k=0$, approaching a discrete Dirac, yielding lower Doppler-domain sidelobes, but the reduced amplitude variability diminishes temporal decorrelation, leading to higher sidelobes in the zero-Doppler cut.
 
\section{Numerical Results}\label{section3}
This section provides simulation results supporting the theoretical analysis.
We evaluate the sensing performance of PSK and QAM constellations using the defined metrics, where we consider $N=64$ subcarriers, unless otherwise specified. The analytical approximations have also been validated with higher lengths.
The PA distortion power affecting the waveforms is varied by changing the IBO at the PA. All the AFs are evaluated in accordance with their average AF performance $\tilde{A}(\tau, \nu)$, i.e. $ \tilde{A}(\tau, \nu) =\frac{1}{N_{MC}}\sum_{m=1}^{N_{MC}}|A^{(m)}(\tau, \nu)|^2$, which experiences $N_{MC}$ runs and $A^{(m)}(\tau, \nu)$ represents the $m^\text{th}$ random realization of AF. In addition, the average AFs are normalized.
\subsection{Zero-Doppler cut Analysis}
\begin{figure}[!t]
  \centering
  \includegraphics[width=1\linewidth]{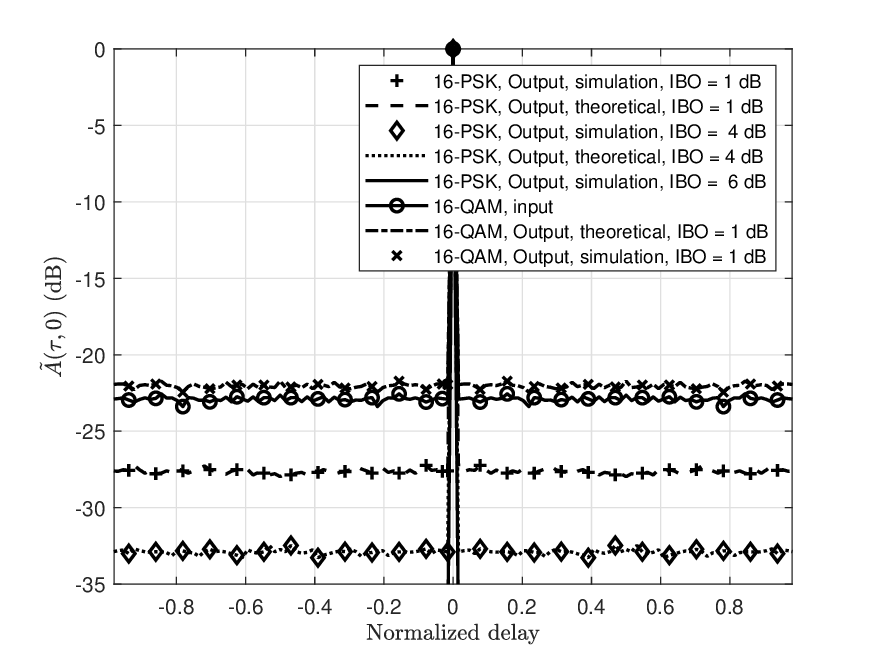}
  \caption{\centering The zero-Doppler cuts for different constellations under CP-OFDM signaling with \& without PA distortions, N = 64. }
      \label{fig11}
\end{figure}
\begin{figure}[!t]
  \centering
  \includegraphics[width=1\linewidth]{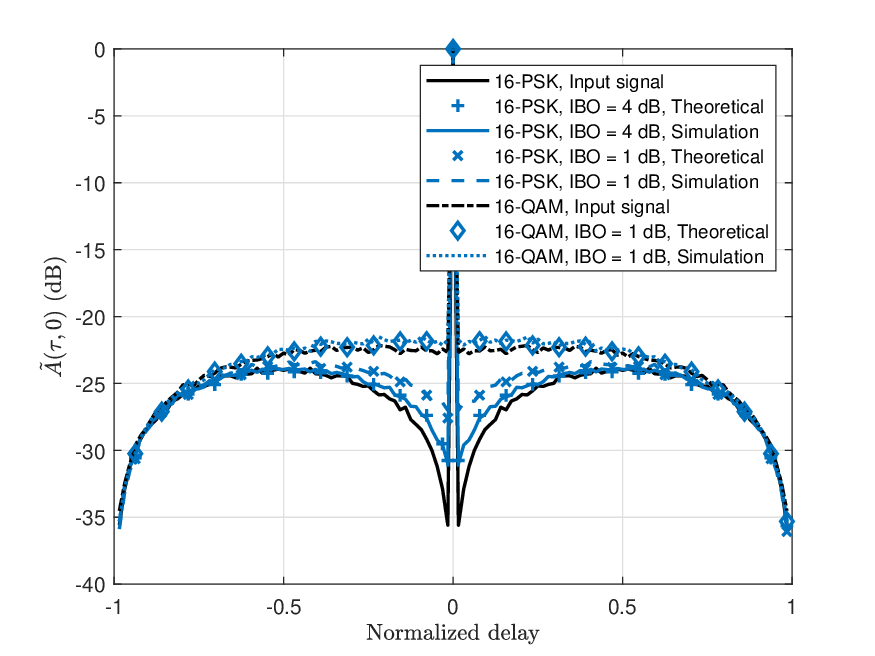}
  \caption{\centering The zero-Doppler cuts for different constellations under OFDM signaling with \& without PA distortions, N = 64.}
\label{fig1}
\end{figure}
The zero-Doppler cuts for 16-PSK and 16-QAM constellations, with and without a CP, are shown in Fig.~\ref{fig11} and Fig.~\ref{fig1}, alongside their analytical counterparts derived from \eqref{A22} and \eqref{SELACF}, respectively. We clearly see that PA distortions elevate the ranging sidelobes. The stronger the distortion, the higher the sidelobe levels, thereby altering the desired AF shape, particularly for PSK constellations. This effect is especially pronounced in the CP-OFDM case in Fig.~\ref{fig11}, where the ideal shape characterized by zero-level sidelobes in the linear scale is severely degraded. For instance, at an IBO = 1 dB, the sidelobe level increases to -27.63 dB, reducing the gap with QAM to about $\approx$4.8 dB.
Interestingly, QAM constellations already exhibit higher sidelobe levels than PSK, but the elevation of these sidelobes under PA distortion is less pronounced in QAM than in PSK, indicating that PSK signals are more vulnerable to PA-induced nonlinearities.\\
\begin{figure}[!t]
  \centering
  \includegraphics[width=1\linewidth]{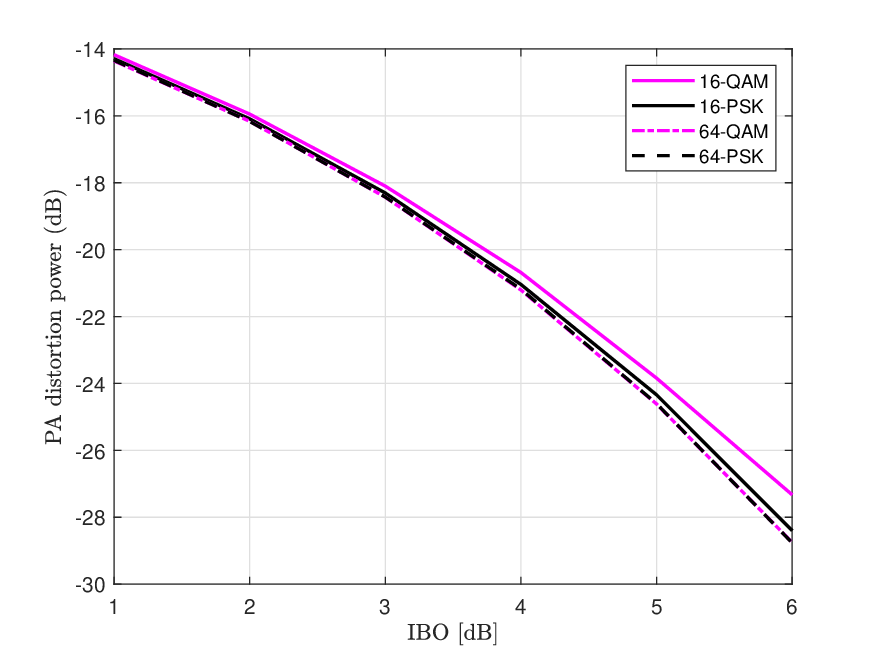}
  \caption{\centering PA distortion power vs IBO, N = 1024.  }
      \label{fig111}
\end{figure}
However, one might intuitively expect QAM to be more affected by clipping due to its amplitude fluctuations. To verify this, we show in Fig.~\ref{fig111} the average PA distortion power obtained from 1000 Monte Carlo realizations of 16/64-PSK and 16/64-QAM constellations with $N=1024$. At low IBO, both PSK and QAM signals are heavily distorted, yielding similar distortion powers. As the IBO increases, PSK, with its constant envelope, experiences minimal clipping, while QAM still has occasional high-amplitude peaks that exceed the SEL saturation threshold, resulting in slightly higher distortion power; however, this difference diminishes for higher-order constellations.

Therefore, the difference in AF behavior between PSK and QAM cannot be primarily attributed to the severity of distortion. Instead, as explained in \hyperref[coroll2]{Corollary .3}, it arises from their intrinsic autocorrelation properties. As shown in Fig.~\ref{fig11} and Fig.~\ref{fig1}, QAM signals already exhibit higher sidelobes due to poor autocorrelation, and thus, the relative increase in the sidelobe level is small. Unlike PSK signals, which exhibit ideal autocorrelation properties in a linear system, these signals suffer more noticeable degradation, resulting in the rise of sidelobes that were previously well-suppressed, especially at low IBO. Consequently, although the use of QAM constellations results in higher sidelobes, the relative sidelobe rise is more pronounced for PSK, making PSK signals more sensitive to nonlinear distortion despite their constant envelope.\\
\begin{figure}[!t]
  \centering
  \includegraphics[width=1\linewidth]{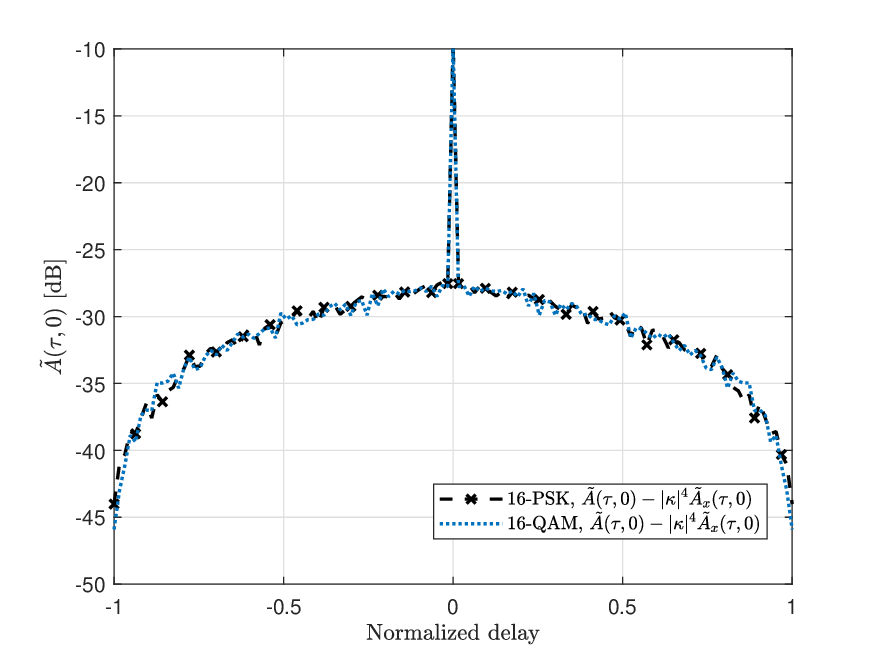}
  \caption{\centering Zero-Doppler cuts with simulated $\tilde{A}(l, 0)$and $\tilde{A}_{x}(l, 0)$, and the analytical $\tilde{A}(l, 0) - |\kappa|^4 \tilde{A}_{x}(l, 0)$, of different constellations under OFDM signaling, N = 64. }
          \label{fig4}
\end{figure}
These observations are further supported by the analytical results in Fig.~\ref{fig4}. We plot $\tilde{A}(l,0)-\tilde{A}_{x}(l,0) $ from \eqref{A22}, which isolates the distortion contribution from the original signal component, normalized to the mainlobe peak of the output signal, to visualize the relative sidelobe rise. The results show that the zero-Doppler cut shape is similar for both PSK and QAM constellations, and that these distortions predominantly affect the lower range bins (i.e., shorter delays), gradually diminishing at larger delays. Hence, in dense-target scenarios, reflections from near weak targets may be masked by the elevated sidelobe floor. Moreover, PA distortion causes the sidelobes of PSK signals to rise toward, but not surpass, those of QAM, since the distortion-related term in Fig.~\ref{fig4} exhibits a similar sidelobe level for both modulations, while QAM originally exhibits higher sidelobes due to modulation-induced characteristics.
\begin{figure}[!t]
  \centering
  \includegraphics[width=1\linewidth]{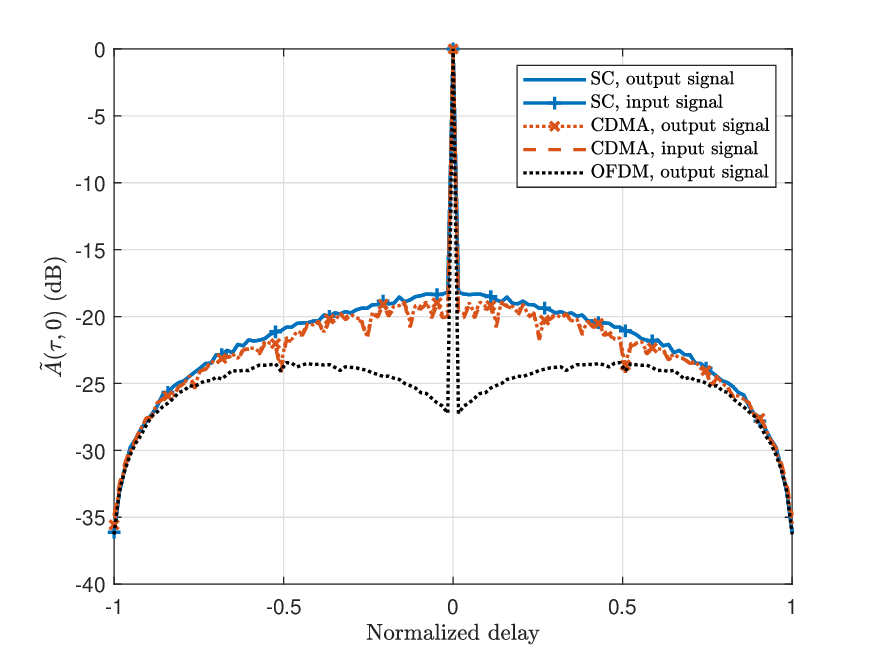}
  \caption{\centering The zero-Doppler cuts of the 16-PSK constellation with PA distortions, under various signaling schemes, N = 64.}
      \label{fig3}
\end{figure}

\begin{figure}[!t]
  \centering
  \includegraphics[width=1\linewidth]{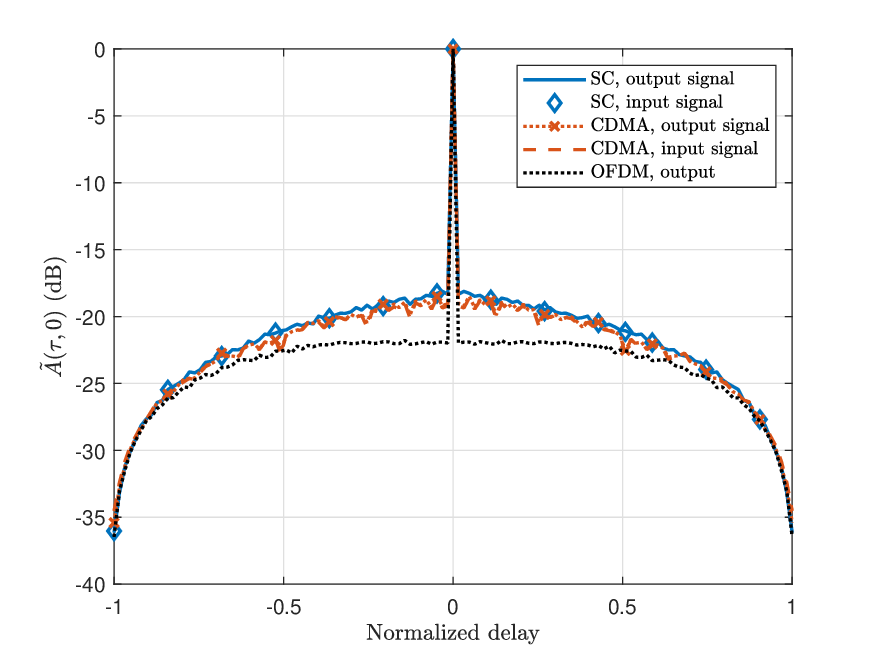}
  \caption{\centering The zero-Doppler cuts of the 16-QAM constellation with PA distortions, under various signaling schemes, N = 64. }
      \label{fig31}
\end{figure}
\subsubsection{Comparison with Classical Signaling Schemes}

To contextualize the performance of OFDM under PA distortions, we next compare it with other classical signaling schemes. In typical communication systems, a vector of $N$ information symbols $\mathbf{s}$ is mapped onto an orthonormal basis in the time domain, represented by a unitary transformation matrix $\mathbf{U}$. The transmit signal can be expressed in a generic form that represents most orthogonal communication signaling schemes as $\mathbf{z} = \mathbf{U} \mathbf{s}$. Specifically, OFDM corresponds to $\mathbf{U} = \mathbf{F}_N$, where $\mathbf{F}_N$ denotes the $N$-point discrete Fourier transform (DFT) matrix; single-carrier (SC) transmission is obtained with $\mathbf{U} = \mathbf{I}_N$; and code-division multiple access (CDMA) can be represented by a Hadamard matrix $\mathbf{U}$ of size $N \times N$.
We illustrate in Fig.~\ref{fig3} and Fig.~\ref{fig31} the zero-Doppler cuts of the 16-PSK and 16-QAM constellations under the aforementioned signaling schemes, for an IBO = 1. The CDMA scheme with Hadamard matrix as a signaling basis performs similarly to the SC approach. Moreover, both exhibit sidelobe levels with no significant increase. In contrast, OFDM achieves superior performance compared to both SC and CDMA for PSK and QAM modulations, with nearly an order of magnitude improvement for PSK. This gain arises primarily from the inherently lower sidelobes of its multicarrier structure, rather than from any degradation in SC or CDMA performance. The results observed for CDMA and SC can also be explained with \hyperref[coroll2]{Corollary .3}.

\subsubsection{Integrated Sidelobe level analysis}
\begin{figure}[!t]
  \centering
  \includegraphics[width=1\linewidth]{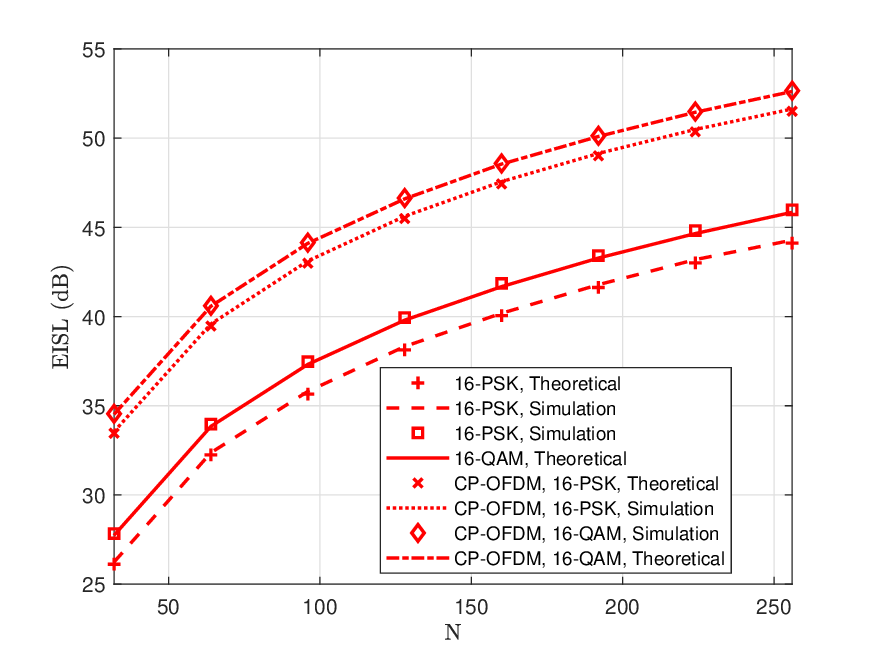}
  \caption{\centering The resultant EISL computed from \eqref{EISLSEL} with varying sequence length, IBO = 1 dB. }
          \label{fig52}
\end{figure}

To show the overall performance of OFDM signals under distortion, we illustrate in Fig.~\ref{fig52} the EISL following \eqref{EISLSEL} with respect to the number of symbol samples $N$, where both cases with and without CP are considered. First of all, the EISL increases with the number of subcarriers conforming to \hyperref[coroll1]{Corollary 1}. Although QAM signals with CP typically exhibit the highest sidelobe levels due to both PA nonlinearity and envelope variation, PSK signals with a CP can still show higher EISL than non-CP QAM signals. This arises because the CP enforces temporal coherence at symbol boundaries, elevating sidelobes at higher delays and thus increasing the overall energy spread in the ambiguity domain.

\begin{figure}[!t]
  \centering
  \includegraphics[width=1\linewidth]{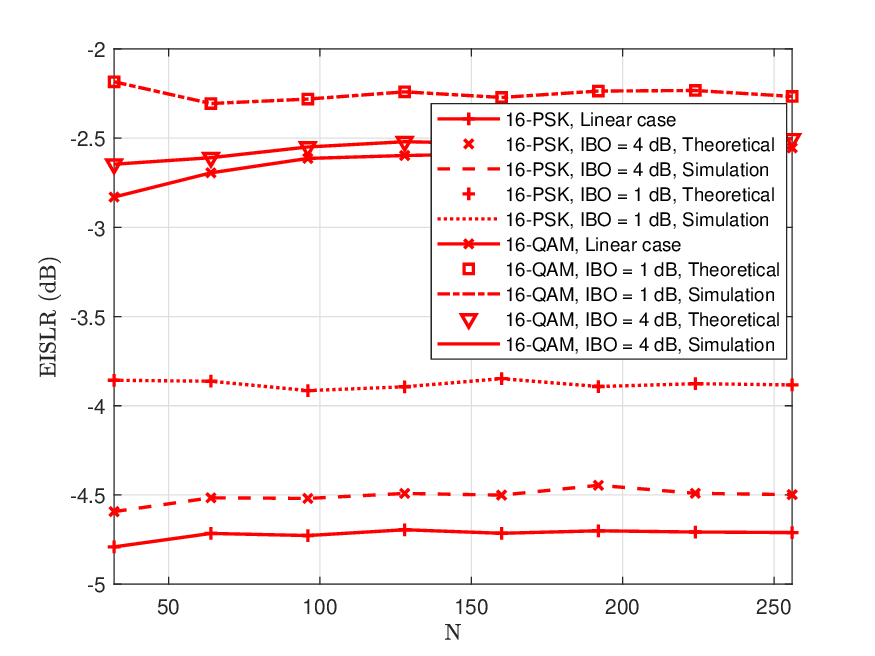}
  \caption{\centering The resultant EISLR computed from \eqref{Buss_EISL} for different constellations under OFDM signaling with varying sequence length. }
          \label{fig51}
\end{figure}

We also illustrate in Fig.~\ref{fig51} the resultant EISLR in the aperiodic case (following \eqref{Buss_EISL}) with respect to the number of symbol samples $N$ for different constellations. We observe that PSK constellations still achieve lower EISLR values compared to QAM. For both modulation schemes, increasing the level of the nonlinear distortion leads to higher sidelobe levels, which was expected from previous results and may also be inferred from the previous analysis. Additionally, the EISLR shows minimal variation as $N$ increases. At first glance, this might suggest that increasing the number of symbols does not mitigate the distortion impact on the sidelobes; however, this conclusion would be misleading, since the EISLR only quantifies the total sidelobe energy and not the sidelobe level at each delay. By the scaling law, while the EISL increases with the sequence length, the mainlobe levels are also on the rise, which has been predicted in our theoretical results.  Therefore, to properly capture the effect of $N$, it is necessary to complement these results with a study of the peak-to-sidelobe level ratio (PSLR), which indicates the ratio between the peak powers of the highest sidelobe and the mainlobe.

\begin{figure}[!t]
  \centering
  \includegraphics[width=1\linewidth]{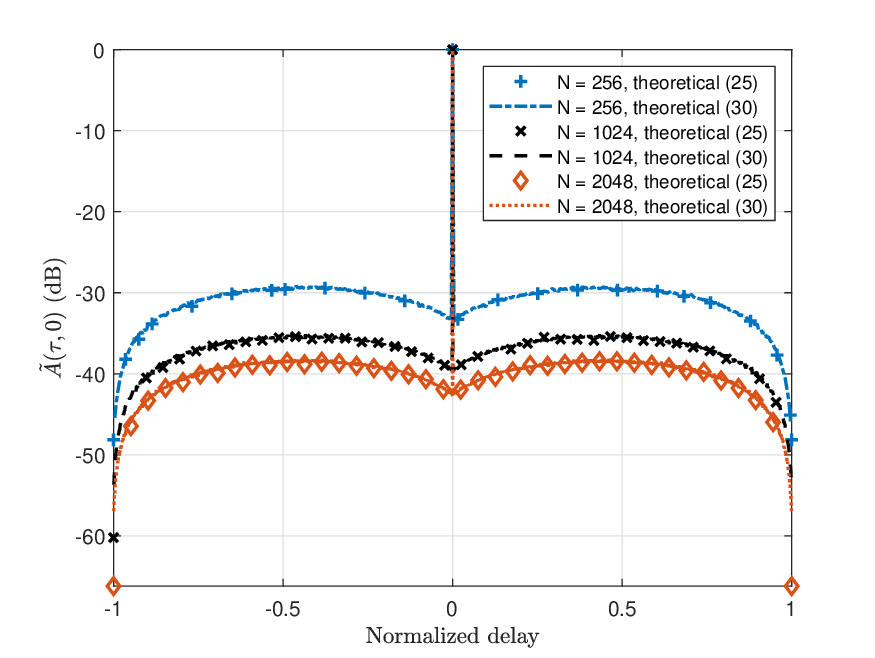}
  \caption{\centering The zero-Doppler cuts for 16-PSK with PA distortions for varying sequence lengths}
      \label{fig1111}
\end{figure} 
We investigate the PSLR of the OFDM signals, where 16-PSK signals are analyzed in Fig.~\ref{fig1111} with varying sequence lengths. As the number of subcarriers increases, the gap between the highest sidelobe level and the mainlobe increases, resulting in lower PSLR values for longer sequences. And a 3 dB performance gain in PSLR may be obtained simply by doubling $N$. This indicates that using longer sequences can enhance the AF shape by reducing the peak sidelobe levels relatively to the mainlobe. Additionally, these longer sequences provide an opportunity to validate the analytical expressions in \eqref{A22} and \eqref{SELACF} under higher sequence lengths.
\begin{figure}[!t]
  \centering
  \includegraphics[width=1\linewidth]{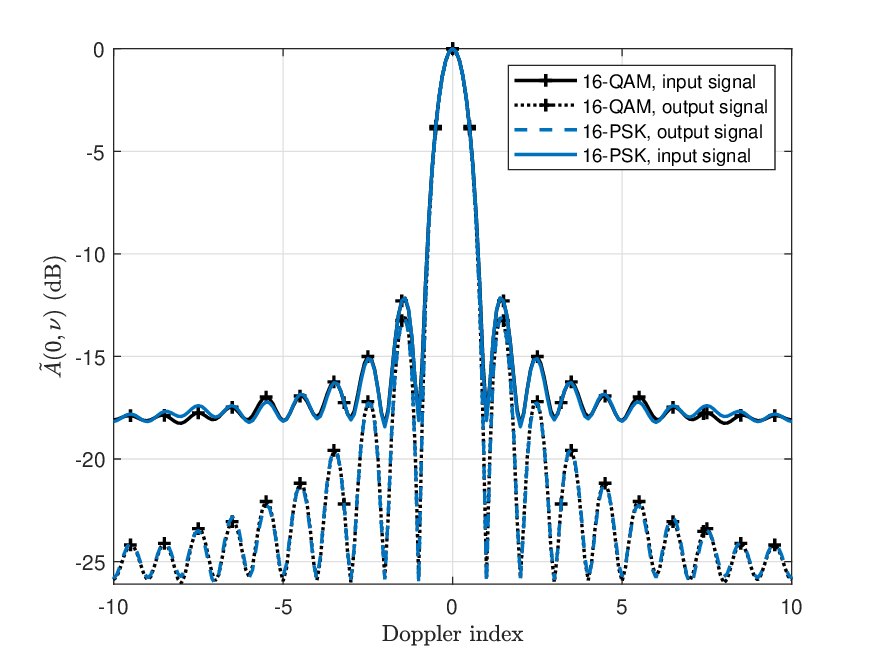}
  \caption{\centering The zero-delay cuts for different constellations under OFDM signaling with \& without PA distortions, N = 64, IBO = 4 dB. }
  \label{zerodelay}
\end{figure}

\begin{figure}[!t]
  \centering
  \includegraphics[width=1\linewidth]{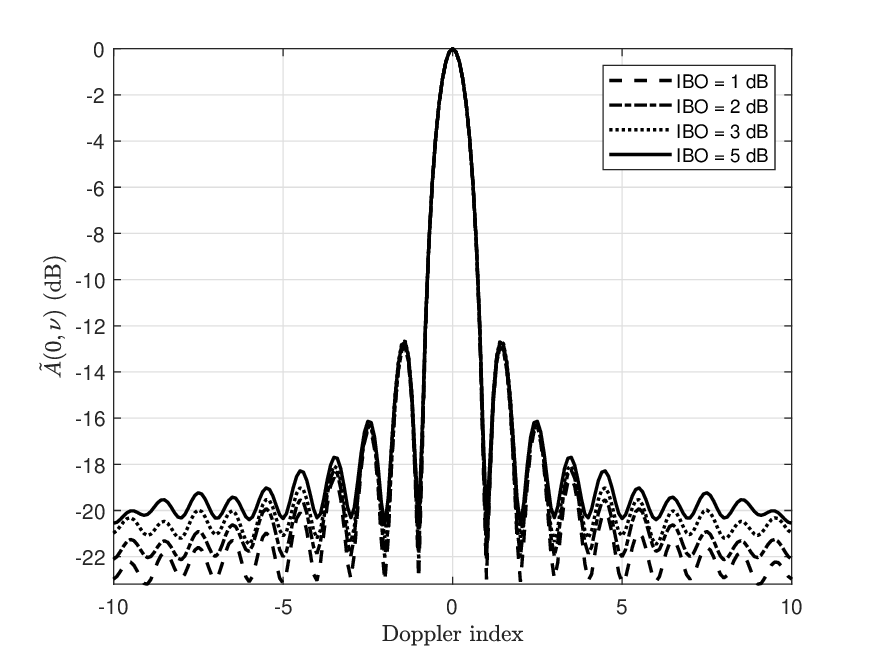}
  \caption{\centering The zero-delay cuts for 16-QAM under OFDM signaling with PA distortions, N = 64.}
  \label{zerodelayQAM}
\end{figure}

\subsection{Zero-delay cut Analysis}
Fig.~\ref{zerodelay} illustrates the zero-delay cut after amplification with IBO = 4 dB. We observe that the sidelobe levels of the output signal become lower, and that both curves of PSK and QAM are nearly identical when we use unit average power, which is consistent with the analytical study. We also showcase in Fig.~\ref{zerodelayQAM} that the greater the distortion, the lower the sidelobe levels become.

\subsection{Ranging performance}
In addition, we further examine the practical ranging performance of random PSK and QAM symbols under PA distortions. To evaluate the impact of the elevated sidelobes, we consider a use case of detecting a weak target in the presence of interference from adjacent targets.

\begin{figure}[!t]
  \centering
    \subfloat[]{%
        \includegraphics[width=0.5\textwidth]{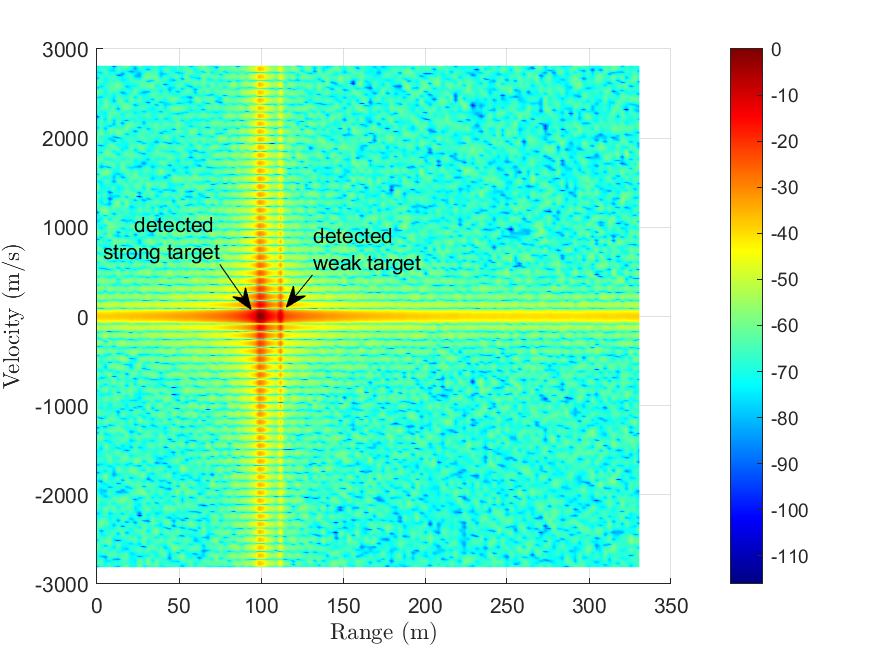}
    }\\
    \subfloat[]{%
        \includegraphics[width=0.5\textwidth]{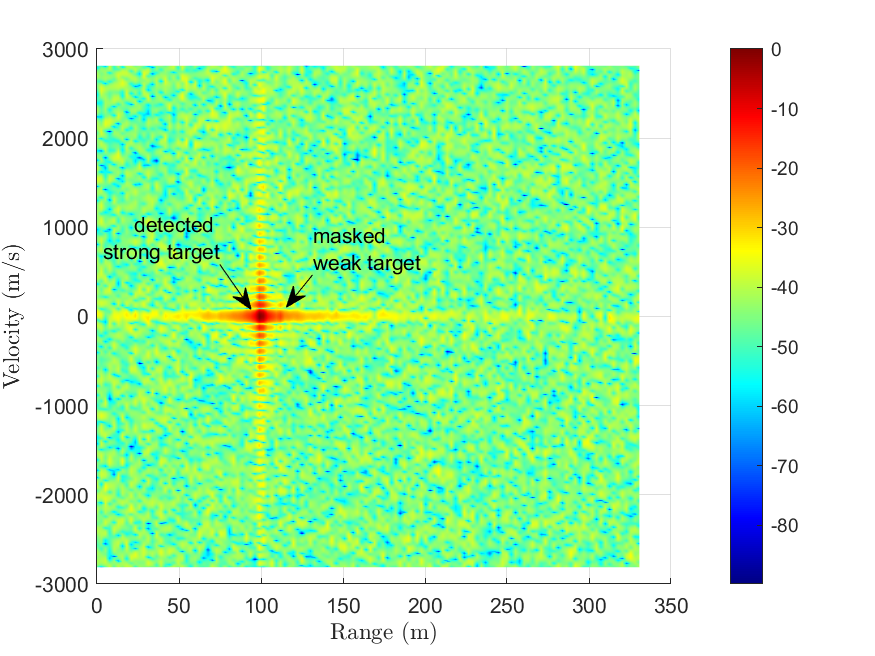}
    }
    \caption{\centering Comparison of (a) linear-case and (b) nonlinear-case periodograms with 16-PSK signals, $N$ = 64, $M$ = 64, IBO = 1 dB, and SNR = 20 dB.}
    \label{fig:RDM}  
\end{figure}

As illustrated in Fig.~\ref{fig:RDM} for a 16-PSK OFDM signal with $N $= 64 subcarriers, an IBO of 1 dB and an SNR of 20 dB, the nonlinear PA introduces spurious components that raise the sidelobe floor of the periodogram, potentially masking weak targets when they fall to the same level as the elevated low-range sidelobes of nearby strong targets. In the linear case, the detection threshold used in radar sensing algorithms is easily set, as a clear gap typically exists between all targets and noise levels. However, due to PA distortions, detection becomes more challenging because the periodogram's noise floor increases. Furthermore, the impact of the division filter is clearly observable under PA distortions, as it further elevates the noise floor compared to the linear case.

To recover the weak target, the smallest of constant false alarm probability (SO-CFAR) detector is exploited \cite{du2023probabilistic}. In particular, we consider a ranging task where a strong target and a weak target need to be simultaneously sensed. By fixing the transmit power to 1, the SNR is defined as the inverse of the noise variance. The probability of false alarm is fixed as~$10^{-4}$ and remains constant.

The two targets are located at the 4$^{th}$ and 8$^{th}$ range cell, respectively, where the reflection power of the first target is 10 dB higher than the one at the 8th range cell. As a result, the latter may be obscured by the noise floor of the periodogram and fall below the SO-CFAR detection threshold.

\begin{figure}[!t]
  \centering
    \subfloat{%
  \includegraphics[width=1\linewidth]{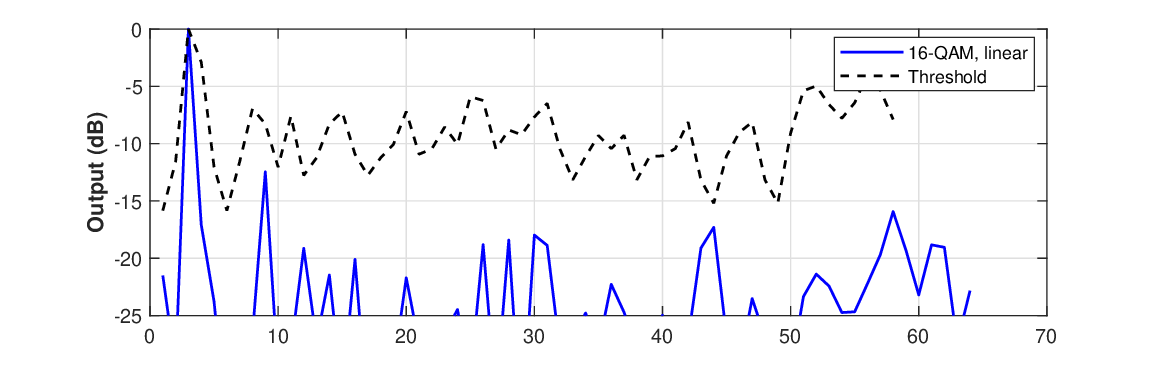}
    }\\
    \subfloat{%
  \includegraphics[width=1\linewidth]{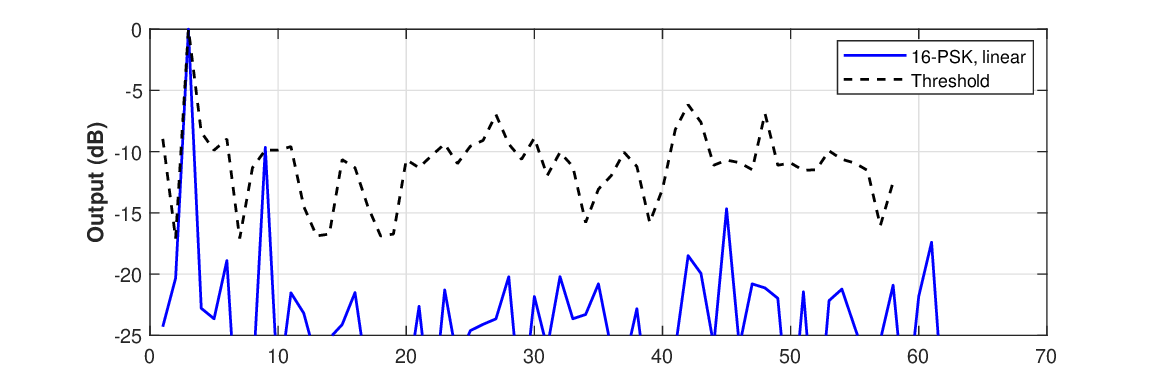}
    }\\
  \subfloat{%
  \includegraphics[width=1\linewidth]{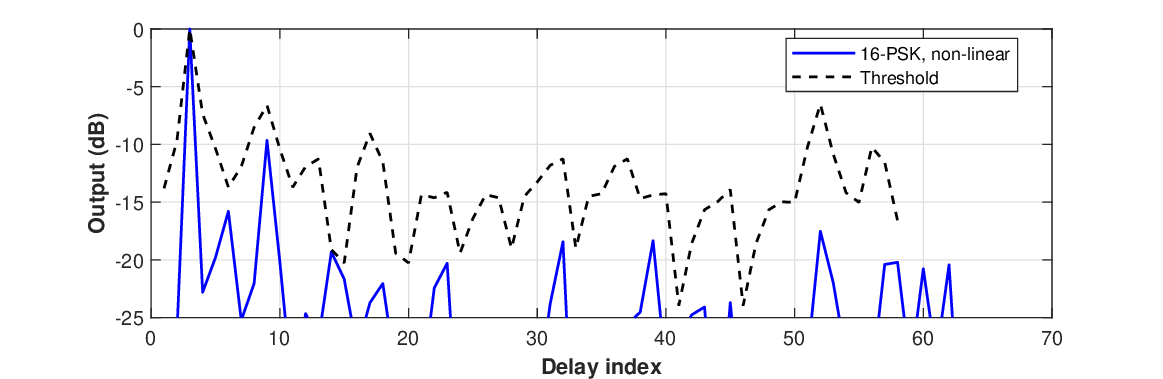}
    }
    \caption{\centering{SO-CFAR results for single realizations of 16-PSK \& 16-QAM, with \& without PA distortions, IBO = 5 dB.}}
    \label{fig5}  
\end{figure}

\begin{figure}[!t]
  \centering
  \includegraphics[width=1\linewidth]{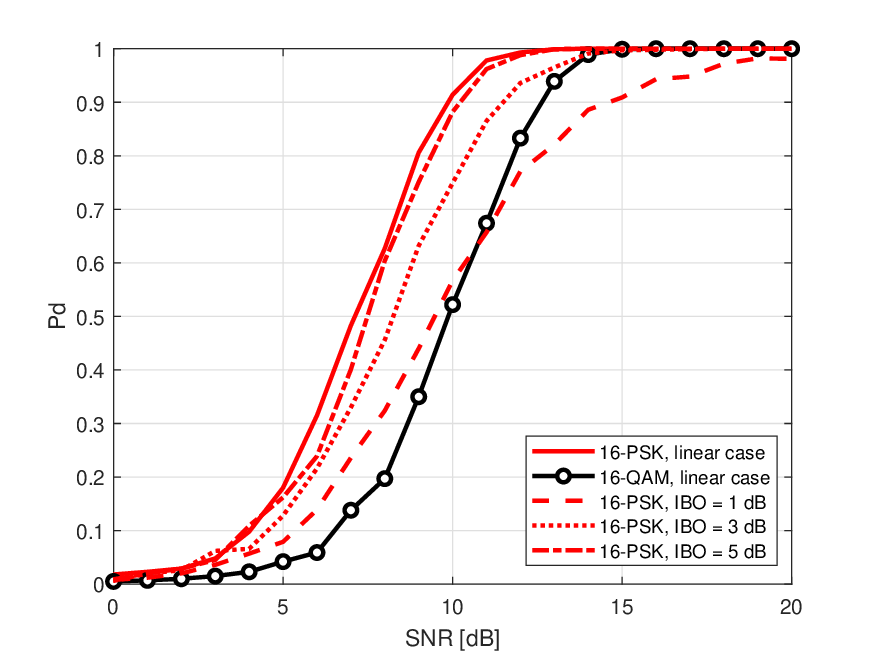}
  \caption{\centering Probability of detection of 16-PSK \& 16-QAM versus sensing SNR, with \& without PA distortions. }
      \label{fig61}
\end{figure}

Fig.~\ref{fig5} illustrates the performance of SO-CFAR for random realizations of 16-QAM and 16-PSK, showcasing that 16-PSK is more efficient for weak target detection due to the lower random sidelobes of the strong target, while in the non-linear case, 16-PSK results in higher sidelobes, thus a higher threshold on top of the weak target. Nevertheless, the results of single realizations are not sufficiently convincing.\\ To evaluate the average performance, Fig.~\ref{fig61} shows the probability of detection ($Pd$) of the weak target versus sensing SNR, averaged over 1000 Monte Carlo trials, for 16-PSK and 16-QAM modulations. In this sense, a lower EISL/EISLR intuitively suggests a better ranging performance. This intuition has been confirmed by our simulation, where the linear case waveform always outperforms the nonlinear case. As the distortion level rises, the $Pd$ of 16-PSK approaches that of 16-QAM. This behavior results from the increased sidelobe floor, which obscures the weak target return and ultimately prevents the detector from correctly identifying it.
Furthermore, for SNR values above 10.5 dB, 16-PSK exhibits lower $Pd$ values than the linear 16-QAM. Although PSK is typically favored for sensing due to its constant-envelope and its ability to preserve noise statistical characteristics after the division filter conducted before the 2D-FFT, this benefit diminishes at high SNR ($\geq$ 10.5 dB). In this regime, the main limitation is no longer noise, but nonlinear distortion from the PA. However, a higher detection accuracy can be achieved ultimately by increasing the SNR.

\begin{figure}[t]
  \centering
  \includegraphics[width=1\linewidth]{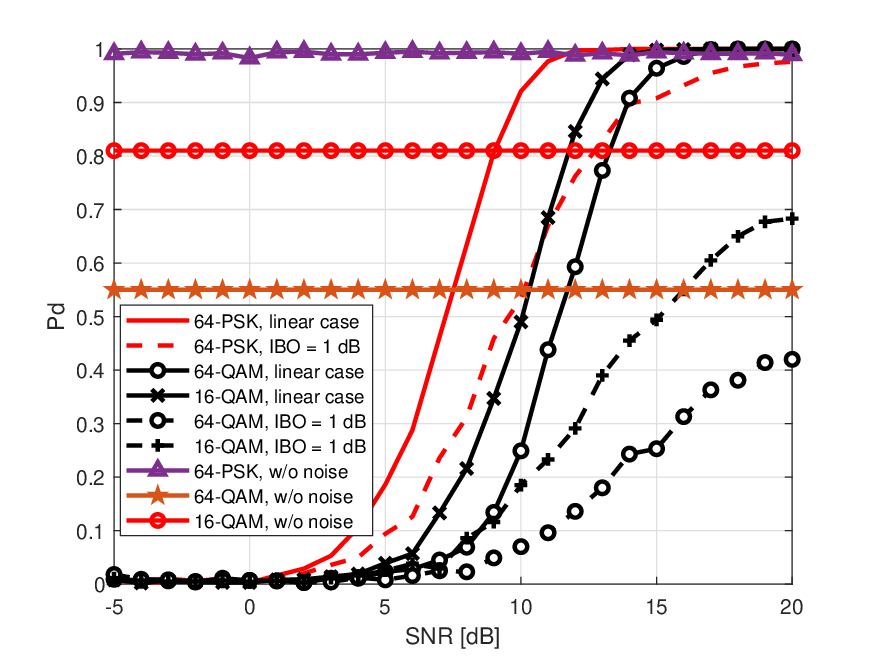}
  \caption{\centering Probability of detection of different constellations versus sensing SNR, with \& without PA distortions, IBO = 1 dB. }
      \label{fig6}
\end{figure}

We further illustrate this effect on QAM constellations in Fig.~\ref{fig6}. QAM constellations exhibit significantly degraded performance in the presence of PA distortions. Notably, the performance deteriorates further with increasing the modulation order.
\\
We also include the distortion-limited scenario, where the noise level is set to zero, and in which case we expect an asymptotic behavior.
The $Pd$ of PSK in this case is always ideal (=1). Whereas, the level of horizontal asymptotes for QAM indicates the presence of a performance ceiling caused by the PA distortions beyond which increasing the SNR does not yield a performance improvement, as predicted in the analytical study. The system behaves as if it were operating at a lower SNR. We remind that the effective SNR ($\text{SNR}_{\text{eff}}$) is the SNR at which a linear system would achieve the same performance as a nonlinear system operating under infinite SNR. This value can be inferred by projecting the performance of the distortion-limited case onto the linear-case curves.
\\
From Fig.~\ref{fig6}, for example, 16-QAM under PA distortion asymptotically achieves an upper limit $Pd$$\approx0.8$, while the linear system reaches the same $Pd$ at $\text{SNR}_{\text{eff}}=12$dB. Similarly, 64-QAM plateau at $Pd$$\approx0.56$, also with an $\text{SNR}_{\text{eff}}=12$dB. 

In addition to the modulation order, we have shown that longer sequence lengths affect the ranging sidelobe levels, hence the probability of detection. For instance, increasing the length of the transmitted serial stream (corresponding to the number of subcarriers $N$ and the frame size $M$) can improve performance in the distortion-limited regime and raise the upper detection limits, especially for QAM symbols. However, it is known that for vehicular communications (e.g., Intelligent Transportation Systems), it is advantageous to keep the packet length quite short \cite{triwinarko2021phy}. So, a trade-off must be considered.

\section{Conclusion}
In this work, we investigated the impact of non-ideal power amplifiers (PAs) on target detection in communication-centric ISAC systems. Our findings reveal a performance ceiling that cannot be overcome merely by increasing the signal power. Furthermore, we analyzed the effect of PA-induced distortion on the ambiguity function (AF) across both the delay and Doppler domains. Analytical derivations under different distortion modeling demonstrate that PA nonlinearities increase the sidelobe levels in the delay domain while reducing them in the Doppler domain. Simulation results using OFDM-based ISAC signals corroborate the analysis, showing that nonlinear artifacts degrade ranging probability and often diminish the sensing advantages of unimodular signaling. These results highlight that the theoretical superiority of constant-modulus waveforms does not necessarily hold under realistic hardware constraints, and that QAM-based signaling can further compromise sensing performance more than usually recognized. Additionally, extending the signal length or increasing the number of transmitted symbols can mitigate the impact of PA distortion by lowering sidelobe levels relatively to the mainlobe. However, such improvement comes at the cost of longer transmission durations, which are not feasible in highly mobile vehicular networks constrained by limited channel coherence time. This motivates us to optimize the AF under practical conditions. Nonetheless, as the AF is solely a waveform sensing metric, it cannot fully capture the end-to-end sensing performance. As conducted in this work, the performance of the detection algorithm must always complement the AF analysis. Therefore, with the PAPR challenge remaining to be tackled, the design of a nonlinearity-robust sensing receiver will be the focus of our future work.

\appendices
\section{Proof of \eqref{Buss_EISL} and \eqref{Buss_EISLR}}\label{A}
The modulo operator is omitted here for notational simplicity, and all discrete-time signals are assumed to be $N$-periodic.
We take the expectation of $|A(l, 0)|^2$ in equation \eqref{A22}
\begin{equation}
\begin{aligned}
    \mathbb{E}\left[ |A(l, 0)|^2 \right] &= |\kappa|^4 \mathbb{E}\left[ |A_x(l, 0)|^2 \right]  + \mathbb{E}\left[ |A_d(l, 0)|^2 \right]\\
 & \quad \,+ 2\Re\{\kappa^2\mathbb{E}\left[ A_x(l,0)A_d^\ast(l,0) \right]\} ,
 \end{aligned}
\label{}
\end{equation}
where
\begin{equation*}
\mathbb{E}\left[ |A_x(l, 0)|^2 \right] = \frac{1}{N} \sum_{\substack{p=0,\\ p'=0}}^{N-1} \mathbb{E}\left[ x(p)x^\ast(p-l)x^\ast(p')x(p'-l) \right].
\label{}
\end{equation*}
We may simplify $\mathbb{E}\left[ x(p)x^\ast(p-l)x^\ast(p')x(p'-l) \right]$
with
\begin{equation*}
\begin{aligned}
&= 
\begin{cases} 
\mathbb{E} [ |x(p)|^2 |x(p-l)|^2 ],& p = p'\, \\
0 & \, \text{otherwise}.
\end{cases}
 \end{aligned}
\end{equation*}
Therefore,
\begin{equation}
\begin{aligned}
\mathbb{E}\left[ |A_x(l, 0)|^2 \right] &= \frac{1}{N} \sum^{N-1}_{p=0} \mathbb{E}\left[  |x(p)|^2 |x(p-l)|^2\right] .
\label{}
 \end{aligned}
\end{equation}
The components $d(n)$ are i.i.d. distributed in OFDM \cite{ismail2024robustness}, so we have similarly
\begin{equation}
\begin{aligned}
    \mathbb{E}\left[ |A_d(l, 0)|^2 \right] &= \frac{1}{N} \sum_{\substack{p=0,\\ p'=0}}^{N-1} \mathbb{E}\left[ d(p)d^\ast(p-l)d^\ast(p')d(p'-l) \right]\\
    &=\frac{1}{N}  \sum^{N-1}_{p=0} \mathbb{E}\left[  |d(p)|^2 |d(p-l)|^2  \right].
\end{aligned}
\end{equation}

Therefore

\begin{equation}
\begin{aligned}
\text{EISL} &= \frac{1}{N} \sum^{N-1}_{\substack{l=1-N,\\ l\neq 0}} \sum^{N-1}_{p=0} |\kappa|^4\mathbb{E}\left[  |x(p)|^2 |x(p-l)|^2  \right]\\ &\qquad \qquad\qquad \qquad+ \mathbb{E}\left[  |d(p)|^2 |d(p-l)|^2  \right]\\  & \quad+ \frac{2\kappa^2}{N}\Re\{\sum^{N-1}_{\substack{p=0,\\ p'= 0}} \mathbb{E}\big[  x(p)x^\ast(p-l) d^\ast(p')d(p'-l) \big]\} .
 \end{aligned}
\end{equation}

Under the assumption of i.i.d. TD samples and the uncorrelated nature of $x(p)$ and $d(p)$, the EISL can be further simplified as
\begin{equation}
\text{EISL} = (2N-2)\left(|\kappa|^4 \sigma^4 + \sigma_d^4 \right) \, .
\end{equation}
This completes the proof of \eqref{Buss_EISL}.

The expected mainlobe level can be derived from the above demonstration as follows

\begin{equation}
\begin{aligned}
    \mathbb{E}\left[ |A(0, 0)|^2 \right] &= \frac{1}{N} \sum^{N-1}_{p=0}|\kappa|^4\mathbb{E}\left[  |x(p)|^4  \right] + \mathbb{E}\left[  |d(p)|^4 \right]\\
& \quad \,+ \frac{2\kappa^2}{N}\Re\{\sum_{\substack{p=0,\\ p'=0}}^{N-1}  \mathbb{E}\big[  |x(p)|^2 |d(p')|^2 \big]\} \\
 &=|\kappa|^4 \mathbb{E}\left[  |x(p)|^4  \right] + \mathbb{E}\left[  |d(p)|^4 \right] \\& \quad+ \frac{2\kappa^2}{N}\sum_{\substack{p=0,\\ p'=0}}^{N-1} \mathbb{E}\big[  |x(p)|^2 |d(p')|^2 \big] ,
 \end{aligned} \label{appendixmainlobe}
\end{equation}

For large $N$, by the central limit theorem, $x(p)$ is approximately complex Gaussian, thus \small$ \mathbb{E}\left[  |x(p)|^4  \right] =  |\kappa|^4 2\big( \mathbb{E}\left[  |x(p)|^2  \right] \big)^2 = |\kappa|^4 2\sigma^4  $ \normalsize, and under the same assumptions used in the EISL final approximation, the expectation of the mainlobe level can be further simplified as,

\begin{equation}
\begin{aligned}
\mathbb{E}\left[ |A(0, 0)|^2 \right] &= 
|\kappa|^4 \mathbb{E}\left[  |x(p)|^4  \right] + \mathbb{E}\left[  |d(p)|^4 \right] \\ & \quad+ \frac{2\kappa^2}{N}N^2  \mathbb{E}\big[  |x(p)|^2 \big] \mathbb{E}\big[ |d(p')|^2 \big] \\ &= |\kappa|^42\sigma^4 + \mathbb{E}\left[  |d(p)|^4 \right]  + 2 \kappa^2N\sigma^2\sigma^2_d.
 \end{aligned} \label{appendixmainlobe}
\end{equation}

This completes the proof of \eqref{Buss_EISLR}.

\section{Proof of the expressions of the joint probabilities in \eqref{SELAF}}\label{C}
To evaluate the probabilities in \eqref{SELAF}, we first consider the straightforward case that occurs when the time shift $l$ extends beyond the duration of one OFDM symbol i.e., $l=\ell N + \Delta l, \ell \in \mathbb{Z} \backslash \{0\}$ and $0\leq \Delta l <N$. In this case, the correlation either involves zero-padding or the subsequent symbol within the transmitted frame. Since OFDM symbols are statistically independent, $x(p)$ and  $x(p-l)$ are uncorrelated. Therefore, the joint probabilities are reduced to the clipping and non-clipping probabilities of $|x(p)|$ (i.e. marginal probabilities) which are given respectively by,
\begin{equation}
\begin{aligned}
\mathbb{P}(|x(p)| > V_{sat}) &= e^{\frac{-V_{sat}^2}{\alpha^2\sigma^2}} = e^{-Y^2} \qquad\quad\,\, , \\
\mathbb{P}(|x(p)|\leq V_{sat}) &= 1 - e^{\frac{-V_{sat}^2}{\alpha^2\sigma^2}} = 1 -  e^{-Y^2}.
\end{aligned}
\end{equation}
where $Y = \frac{V_{sat}}{\alpha\sigma} $.\\

Another straightforward case is when $|x(p)|$ and  $|x(p-l)|$ are fully correlated i.e. $ l = 0$,
\begin{align}
\mathbb{P}(|x(p)| \leq V_{sat},\ |x(p-l)| \leq V_{sat}) &=\mathbb{P}(|x(p)| \leq V_{sat}) \nonumber \\&= 1 - e^{-Y^2}.
\end{align}
When the time shift $l \in ]0, N]$, the samples $x(p)$ and $x(p-l)$ are generally statistically correlated. Therefore, the probabilities in \eqref{SELACF} are each the joint probabilities of statistically dependent variables.
We remind that due to the central limit theorem, $x(p)$ and $x(p-l)$ are jointly circular complex Gaussian random variables. Consequently, the magnitudes $|x(p)|$ and $|x(p-l)|$ follow a bivariate Rayleigh distribution with the normalized correlation coefficient defined as 
\small
\begin{equation}
\rho =
\frac{\operatorname{Cov}\!\big(|x(p)|^2,|x(p-\ell)|^2\big)}
{\sqrt{\operatorname{Var}(|x(p)|^2)\,\operatorname{Var}(|x(p-\ell)|^2)}} .
\end{equation}
\normalsize
When $\rho(l) \in ]0, 1 ]$, the joint probabilities in \eqref{SELACF} can be obtained using the closed form of the cumulative distribution function (CDF) of bivariate Rayleigh distributions in \cite{simon2004digital}. We denote by $\mathbb{P}_<(Y)$ the joint probability that both magnitudes are below the clipping threshold. This probability is defined in \eqref{approxi} for $\rho(l) \in ]0, 1]$, i.e., $0< l\leq N$, with the appropriate changes in notation,
\begin{align}
\mathbb{P}_<(Y) &=\mathbb{P}(|x(p)| \leq V_{sat},\ |x(p-l)| \leq V_{sat}) \nonumber\\ &= 1 - 2e^{-Y^2} + \frac{1}{2\pi} \int_{-\pi}^{\pi} e^{ -Y^2 \left[ 
1 +  \frac{(1-\rho(l))}
{1 + 2\sqrt{\rho(l)} \sin \theta + \rho(l)}
\right]} d\theta \,\,.
\label{approxi}
\end{align}
We note that when $ l = N $, the joint probability in equations \eqref{approxi} is consistent with the expected expression with uncorrelated $|x(p)|$ and $|x(p-l)|$,
\begin{equation}
\mathbb{P}_<(Y) = 1 - 2e^{-Y^2} + e^{-2Y^2} =\left(1 - e^{-Y^2}\right)^2.
\end{equation}
The remaining probabilities follow by subtracting from the marginal probabilities as,

\begin{align*}
\mathbb{P}(|x(p)| \leq V_{sat},\, |x(p-l)| > V_{sat}) &= 1 - e^{-Y^2} - \mathbb{P}_<(Y) \, \, , \nonumber \\
 \mathbb{P}(|x(p)| > V_{sat},\, |x(p-l)| \leq V_{sat}) &= 1 - e^{-Y^2} - \mathbb{P}_<(Y) \, , \nonumber\\
\mathbb{P}(|x(p)| > V_{sat},\, |x(p-l)| > V_{sat}) &= 2e^{-Y^2} + \mathbb{P}_<(Y) -1.
\end{align*}

\bibliographystyle{IEEEtran}
\bibliography{main.bib}

@book{simon2004digital,
  title={{Digital communication over fading channels}},
  author={Simon, Marvin K and Alouini, Mohamed-Slim},
  year={2004},
  publisher={John Wiley \& Sons}
}

@article{varshney2023low,
  title={{Low-PAPR OFDM waveform design for radar and communication systems}},
  author={Varshney, Piyush and Babu, Prabhu and Stoica, Petre},
  journal={IEEE Transactions on Radar Systems},
  volume={1},
  pages={69--74},
  year={2023},
  publisher={IEEE}
}

@phdthesis{braun2014ofdm,
  title={{OFDM radar algorithms in mobile communication networks}},
  author={Braun, Klaus Martin},
  year={2014},
  school={Karlsruhe, Karlsruher Institut f{\"u}r Technologie (KIT), Diss., 2014}
}

@article{rexhepi2024tone,
  title={{Tone Reservation-Based PAPR Reduction Using Manifold Optimization for OFDM-ISAC Systems}},
  author={Rexhepi, Getuar and Ranasinghe, Kuranage Roche Rayan and de Abreu, Giuseppe Thadeu Freitas and others},
  journal={arXiv preprint arXiv:2409.16121},
  year={2024}
}

@article{liu2024ofdm,
  title={{OFDM achieves the lowest ranging sidelobe under random ISAC signaling}},
  author={Liu, Fan and Zhang, Ying and Xiong, Yifeng and Li, Shuangyang and Yuan, Weijie and Gao, Feifei and Jin, Shi and Caire, Giuseppe},
  journal={arXiv preprint arXiv:2407.06691},
  year={2024}
}

@inproceedings{du2023probabilistic,
  title={{Probabilistic constellation shaping for OFDM-based ISAC signaling}},
  author={Du, Zhen and Liu, Fan and Xiong, Yifeng and Han, Tony Xiao and Yuan, Weijie and Cui, Yuanhao and Yao, Changhua and Eldar, Yonina C},
  booktitle={2023 IEEE Globecom Workshops (GC Wkshps)},
  pages={509--514},
  year={2023},
  organization={IEEE}
}

@inproceedings{xu2024experimental,
  title={{An Experimental Validation of ISAC with Probabilistic Constellation Shaping Under OFDM Signaling}},
  author={Xu, Jingjing and Du, Zhen and Wang, Jie and Xu, Yao},
  booktitle={2024 IEEE International Conference on Unmanned Systems (ICUS)},
  pages={1579--1584},
  year={2024},
  organization={IEEE}
}

@inproceedings{ismail2024robustness,
  title={{Robustness of ISAC Waveforms to Power Amplifier Distortion}},
  author={Ismail, Abdur Rahman Mohamed and Guenach, Mamoun and Sakhnini, Adham and Bourdouk, Andr{\'e} and Steendam, Heidi},
  booktitle={2024 IEEE 4th International Symposium on Joint Communications \& Sensing (JC\&S)},
  pages={1--6},
  year={2024},
  organization={IEEE}
}

@inproceedings{akca2024integrated,
  title={{Integrated Sensing and Communication with Power Amplifier Impairment}},
  author={Akca, Huseyin and Memi{\c{s}}o{\u{g}}lu, Ebubekir and {\c{C}}{\i}rpan, Hakan Ali and Arslan, Huseyin},
  booktitle={2024 6th International Conference on Communications, Signal Processing, and their Applications (ICCSPA)},
  pages={1--6},
  year={2024},
  organization={IEEE}
}

@article{wymeersch2025cross,
  title={{Cross-layer Integrated Sensing and Communication: A Joint Industrial and Academic Perspective}},
  author={Wymeersch, Henk and Tervo, Nuutti and W{\"a}nstedt, Stefan and Saleh, Sharief and Ahlendorf, Joerg and Akgul, Ozgur and Tsekenis, Vasileios and Barmpounakis, Sokratis and Bai, Liping and Beale, Martin and others},
  journal={arXiv preprint arXiv:2505.10933},
  year={2025}
}

@inproceedings{feng2024analysis,
  title={{Analysis of Non-linear Power Amplifier Effect and Digital Predistortion on OFDM Radar}},
  author={Feng, Ruoyu and Bauduin, Marc and Bourdoux, Andr{\'e}},
  booktitle={2024 21st European Radar Conference (EuRAD)},
  pages={248--251},
  year={2024},
  organization={IEEE}
}

@article{bouhadda2014theoretical,
  title={{Theoretical analysis of BER performance of nonlinearly amplified FBMC/OQAM and OFDM signals}},
  author={Bouhadda, Hanen and Shaiek, Hmaied and Roviras, Daniel and Zayani, Rafik and Medjahdi, Yahia and Bouallegue, Ridha},
  journal={EURASIP Journal on Advances in Signal Processing},
  volume={2014},
  number={1},
  pages={60},
  year={2014},
  publisher={Springer}
}

@article{yang2024constellation,
  title={{Constellation design for integrated sensing and communication with random waveforms}},
  author={Yang, Xiaobo and Zhang, Ruonan and Zhai, Daosen and Liu, Fan and Du, Rui and Han, Tony Xiao},
  journal={IEEE Transactions on Wireless Communications},
  year={2024},
  publisher={IEEE}
}

@article{triwinarko2021phy,
  title={PHY layer enhancements for next generation V2X communication},
  author={Triwinarko, Andy and Dayoub, Iyad and Cherkaoui, Soumaya},
  journal={Vehicular Communications},
  volume={32},
  pages={100385},
  year={2021},
  publisher={Elsevier}
}

@article{dapa2023vehicular,
  title={Vehicular communications over OFDM radar sensing in the 77 GHz mmWave band},
  author={Dapa, KB Serge Angelo and Point, Guillaume and Bensator, Saleh and Boukour, Fouzia Elbahhar},
  journal={IEEE Access},
  volume={11},
  pages={4821--4829},
  year={2023},
  publisher={IEEE}
}

@article{dapa2025parametrizations,
  title={Parametrizations of a 77 GHz OFDM joint radar communication},
  author={Dapa, KB Serge Angelo and Elbahhar, Fouzia and Point, Guillaume and Bensator, Saleh},
  journal={IEEE Access},
  year={2025},
  publisher={IEEE}
}

@article{tong2025integrated,
  title={Integrated Sensing, Communication, and Positioning in Cellular Vehicular Networks},
  author={Tong, Xin and Zhang, Zhaoyang and Yang, Yuzhi and Ge, Yu and Yang, Zhaohui and Wymeersch, Henk and Debbah, M{\'e}rouane},
  journal={IEEE Transactions on Vehicular Technology},
  year={2025},
  publisher={IEEE}
}

@inproceedings{wang2018joint,
  title={A joint radar and communication system based on commercially available FMCW radar},
  author={Wang, Chang-Heng and Altintas, Onur},
  booktitle={2018 IEEE Vehicular Networking Conference (VNC)},
  pages={1--2},
  year={2018},
  organization={IEEE}
}

@article{keskin2025fundamental,
  title={Fundamental trade-offs in monostatic ISAC: A holistic investigation towards 6G},
  author={Keskin, Musa Furkan and Mojahedian, Mohammad Mahdi and Lacruz, Jesus O and Marcus, Carina and Eriksson, Olof and Giorgetti, Andrea and Widmer, Joerg and Wymeersch, Henk},
  journal={IEEE Transactions on Wireless Communications},
  year={2025},
  publisher={IEEE}
}

@article{salman2024sensing,
  title={When are sensing symbols required for ISAC?},
  author={Salman, Murat Babek and Demir, {\"O}zlem Tu{\u{g}}fe and Bj{\"o}rnson, Emil},
  journal={IEEE Transactions on Vehicular Technology},
  volume={73},
  number={10},
  pages={15709--15714},
  year={2024},
  publisher={IEEE}
}

@article{nataraja2024integrated,
  title={Integrated sensing and communication (isac) for vehicles: Bistatic radar with 5g-nr signals},
  author={Nataraja, Nikhil K and Sharma, Sudhanshu and Ali, Kamran and Bai, Fan and Wang, Rui and Molisch, Andreas F},
  journal={IEEE Transactions on Vehicular Technology},
  year={2024},
  publisher={IEEE}
}

@inproceedings{temiz2023radar,
  title={Radar-centric ISAC through index modulation: Over-the-air experimentation and trade-offs},
  author={Temiz, Murat and Peters, Nial J and Horne, Colin and Ritchie, Matthew A and Masouros, Christos},
  booktitle={2023 IEEE Radar Conference (RadarConf23)},
  pages={1--6},
  year={2023},
  organization={IEEE}
}

@article{ma2021spatial,
  title={Spatial modulation for joint radar-communications systems: Design, analysis, and hardware prototype},
  author={Ma, Dingyou and Shlezinger, Nir and Huang, Tianyao and Shavit, Yariv and Namer, Moshe and Liu, Yimin and Eldar, Yonina C},
  journal={IEEE Transactions on Vehicular Technology},
  volume={70},
  number={3},
  pages={2283--2298},
  year={2021},
  publisher={IEEE}
}

@article{muja2024real,
  title={Real-time interference mitigation in automotive radars using the short-time fourier transform and L-statistics},
  author={Muja, Robert and Anghel, Andrei and Cacoveanu, Remus and Ciochina, Silviu},
  journal={IEEE Transactions on Vehicular Technology},
  volume={73},
  number={10},
  pages={14617--14632},
  year={2024},
  publisher={IEEE}
}

@article{yang2020dual,
  title={Dual-use signal design for radar and communication via ambiguity function sidelobe control},
  author={Yang, Jing and Cui, Guolong and Yu, Xianxiang and Kong, Lingjiang},
  journal={IEEE Transactions on Vehicular Technology},
  volume={69},
  number={9},
  pages={9781--9794},
  year={2020},
  publisher={IEEE}
}

@inproceedings{dokhanchi2018ofdm,
  title={OFDM-based automotive joint radar-communication system},
  author={Dokhanchi, Sayed Hossein and Shankar, MR Bhavani and Stifter, Thomas and Ottersten, Bj{\"o}rn},
  booktitle={2018 IEEE radar conference (RadarConf18)},
  pages={0902--0907},
  year={2018},
  organization={IEEE}
}

@article{knill2021coded,
  title={Coded OFDM waveforms for MIMO radars},
  author={Knill, Christina and Embacher, Felix and Schweizer, Benedikt and Stephany, Simon and Waldschmidt, Christian},
  journal={IEEE Transactions on Vehicular Technology},
  volume={70},
  number={9},
  pages={8769--8780},
  year={2021},
  publisher={IEEE}
}

@article{bazzi2023integrated,
  title={On integrated sensing and communication waveforms with tunable PAPR},
  author={Bazzi, Ahmad and Chafii, Marwa},
  journal={IEEE Transactions on Wireless Communications},
  volume={22},
  number={11},
  pages={7345--7360},
  year={2023},
  publisher={IEEE}
}

@inproceedings{gourar2025analysis,
  title={Analysis of PAPR-Aware OFDM Waveforms for Integrated Sensing and Communication},
  author={Gourar, Eya and Medjahdi, Yahia and Clavier, Laurent and Gizzini, Abdul Karim},
  booktitle={2025 IEEE 101st Vehicular Technology Conference (VTC2025-Spring)},
  pages={1--6},
  year={2025},
  organization={IEEE}
}

@article{he2024nonlinear,
  title={Nonlinear self-interference cancellation in vehicle networks for full-duplex integrated sensing and communication},
  author={He, Yimin and Zhao, Hongzhi and Shao, Shihai},
  journal={IEEE Transactions on Vehicular Technology},
  volume={73},
  number={9},
  pages={13980--13985},
  year={2024},
  publisher={IEEE}
}

\end{document}